\documentclass{emulateapj}
\usepackage{natbib}
\usepackage{amssymb,amsmath} 

\shorttitle{Very High Energy Emission and Cascade Radiation of GRB Afterglows}
\shortauthors{Huang et al.}
\begin{document}

\title{Very High Energy Emission and Cascade Radiation of Gamma-Ray Burst Afterglows: Homogeneous Versus Wind External Media}
\author{Xiao-Li Huang\altaffilmark{1,3,*}, Ze-Rui Wang\altaffilmark{1,3,*}, Ruo-Yu Liu\altaffilmark{1,3}, Xiang-Yu Wang\altaffilmark{1,3}, En-Wei Liang\altaffilmark{2}}
\altaffiltext{1}{School of Astronomy and Space Science, Nanjing University, Nanjing 210023, China; xywang@nju.edu.cn}
\altaffiltext{2}{Guangxi Key Laboratory for Relativistic Astrophysics, School of Physics Science and Technology, Guangxi University, Nanning 530004, China}
\altaffiltext{3}{Key laboratory of Modern Astronomy and Astrophysics (Nanjing University), Ministry of Education, Nanjing 210023, China}
\altaffiltext{*}{These authors contributed equally to this work.}

\begin{abstract}
Recent detection of sub-TeV emission from gamma-ray bursts (GRBs) represents a breakthrough in the GRB study.
The multi-wavelength data of the afterglows of GRB 190114C support the synchrotron self-Compton (SSC) origin for its sub-TeV emission. We present a comparative analysis on the SSC emission of GRB afterglows in the homogeneous and wind environment in the framework of the forward shock model. The $\gamma\gamma$ absorption of very high-energy photons due to pair production within the source and the Klein-Nishina effect on the inverse-Compton scattering are considered. Generally a higher SSC flux is expected for a larger circum-burst density due to a larger Compton parameter, but meanwhile the internal $\gamma\gamma$ absorption is more severer for sub-TeV emission. The flux ratio between the SSC component and the synchrotron component decreases more quickly with time in the wind medium case than that in the homogenous-density medium case. The light curves of the SSC emission are also different for the two types of media. We also calculate the cascade emission resulted from the absorbed high-energy photons. In the ISM environment with $n\ga 1\,\rm cm^{-3}$, the cascade synchrotron emission could be comparable to the synchrotron emission of the primary electrons in the optical band, which may flatten the optical afterglow light curve at early time ($t<1$ h). In the wind medium with $A_{\ast}\ga 0.1$, the cascade emission in the eV-GeV band is comparable or even larger than the emission of the primary electrons at early time.
\end{abstract}

\keywords{gamma-ray bursts; relativistic jets; non-thermal radiation}

\section{INTRODUCTION}
It was proposed that high-energy afterglows ($>$ 100 MeV) may result from the synchrotron radiations of the shock-accelerated electrons (e.g.,Kumar \& Barniol Duran 2009; Ghisellini et al. 2010; Wang et al. 2010), but the limit of a maximum synchrotron photon energy of about $50\Gamma/(1+z)$ MeV for a burst at redshift of $z$ makes it difficult to explain the observed $\ga 10$ GeV gamma-rays by {\em  Fermi} Large Area Telescope (LAT) at the time when the bulk Lorentz factor $\Gamma$ of the jet has decreased significantly (e.g., Piran \& Nakar 2010). These $>$10 GeV photons could then be produced by synchrotron self-Compton (SSC) emission in the afterglow shocks, which is supported by
multi-band modeling of some LAT-detected GRBs (Wang et al. 2013), particulary the very bright GRB 130427A (Tam et al. 2013; Liu et al. 2013; Ackermann et al. 2014; Fraija et al. 2016). Indeed, afterglow SSC emission has been long predicted to be able to produce high-energy photons (e.g., M\'{e}sz\'{a}ros \&
Rees 1993; Waxman 1997; Chiang \& Dermer 1999; Panaitescu \& Kumar 2000; Sari \& Esin 2001; Wang et al. 2001; Zhang \&
M\'{e}sz\'{a}ros 2001; Granot \& Guetta 2003; Fan et al.2008; Beniamini et al. 2015).

The multi-wavelength data of GRB 190114C strongly support that sub-TeV photons are dominated by the synchrotron self-Compton (SSC) process (MAGIC Collaboration et al. 2019; Derishev \& Piran 2019; Wang et al. 2019; Fraija et al. 2019c). The sub-TeV emission from GRB 180720B (Abdalla et al. 2019) can also be represented with the SSC model (Wang et al. 2019; Fraija et al. 2019b). The SSC emission is sensitive to the density of the external medium, which could be a homogeneous external medium or a stratified wind medium. In this work, we will investigate the differences in the SSC emission of afterglows arising from the two types of external media.

High-energy photons may be absorbed via the pair production process ($\gamma\gamma\rightarrow e^{+}e^{-}$ ) within the source, and the  secondary $e^{+}e^{-}$ pairs could produce cascade emission via the  synchrotron radiation and inverse Compton processes. The $\gamma\gamma$ absorption and pair cascade process have been widely studied in both blazars (Aharonian et al. 2008; Zacharopoulou et al. 2011; Yan \& Zhang 2015) and the prompt emission of GRBs (Pe'er \& Waxman 2005; Gill \& Granot 2018). We here consider this effect in the afterglow phase of GRBs.

This paper is organized as follows. We compare the spectral energy distributions (SEDs) and light curves in two types of media in \S 2. The analyses of cascade radiation initiated by internal $\gamma\gamma$ pair production are presented in \S 3. Conclusions and discussions are presented in \S 4.

\section{Broadband SEDs and Light curves of GRB Afterglows in the Homogeneous and Wind Medium}
Employing the standard dynamic evolution model for GRB afterglows (e.g., Huang et al 1999),  we derive the SEDs and light curves of the GRB afterglows by considering the afterglow emission is produced by electrons accelerated in the forward shocks expanding into the external medium. Two types of media were extensively studied, i.e., the homogeneous medium with a constant density ($n=n_0$; Sari et al. 1998) and the wind medium with density profile as $n(r)=A\,r^{-2}$, where $A = \frac{\dot{M}}{4\pi m_{p}V}=3.0\times10^{35}\,A_{\ast}\,\rm cm^{-1}$ with $A_{\ast}=\frac{\dot{M}/10^{-5}M_{\odot}\,\rm yr^{-1}}{v/10^{3}\,\rm km\,s^{-1}}$,  $\dot{M}$ is the mass-loss rate of the massive star, and $V$ is the constant wind speed for a Wolf-Rayet star (Dai \& Lu 1998; Chevalier \& Li 2000; Panaitescu \& Kumar 2000). The radiation mechanisms are the synchrotron radiations and SSC process of the electrons accelerated in the shocks (e.g., Sari et al. 1998; Sair \& Esin 2001). The distribution of the radiating electrons is taken as a single power-law function $\rm dN/\rm d\gamma_e\propto\gamma_e^{-p}$, where $\gamma_e$ is the electron Lorentz factor and $p$ is the electron spectral index. The synchrotron spectrum is characterized by several power-law segments with breaks at the synchrotron-self-absorbtion frequency ($\nu_a$), the photon frequency from the injected minimum-energy electrons ($\nu_m$), and the cooling photon frequency $\nu_c$. In addition, the SSC component can be calculated by the synchrotron spectrum and the Compton parameter ($Y$ parameter) with break frequencies at $\nu_{a}^{\rm IC}$, $\nu_{m}^{\rm IC}$ and $\nu_{c}^{\rm IC}$ (Panaitescu \& Kumar 2000; Sair \& Esin 2001).

The cross section for the IC scattering is suppressed when the photon energy in the electron rest frame exceeds $\sim m_{e}c^{2}$, which is the so-called Klein-Nishina (KN) effect. This effect is important at sufficiently high energies for GRB afterglows (Nakar et al. 2009; Wang et al. 2010).  The Compton parameter $Y(\gamma_{e})$ depends on the energy of electrons $\gamma_{e}$, which is given by
\begin{equation}
\begin{array}{lll}
Y(\gamma_{e})=\frac{U_{\rm syn}[\nu<\nu_{\rm KN}(\gamma_{e})]}{U_{B}},
\end{array}
\end{equation}
where $\nu_{\rm KN}$ is the critical frequency of scattering
photons above which the scatterings with electrons of energy $\gamma_e$
just enter the KN scattering regime, $U_{\rm syn}[\nu<\nu_{\rm KN}(\gamma_{e})]$ is the energy density
of the synchrotron photons with frequency below $\nu_{\rm KN}$, and $U_{\rm B}$ is the energy density of the magnetic field.

The $Y(\gamma_{e})$ parameter affects the electron radiative cooling function and modify the electron distribution. The modified electron distribution in the fast cooling case is given by
\begin{equation}
N(\gamma_{e})=\frac{C_{1}}{1+Y(\gamma_{e})}\left\{
\begin{array}{lll}
\gamma_{e}^{-2},&& \gamma_{c}<\gamma_{e}<\gamma_{m},\\
\gamma_{m}^{p-1}\gamma_{e}^{-p-1},&& \gamma_{m}<\gamma_{e},
\end{array}\right.
\end{equation}
and in the slow cooling case,
\begin{equation}
N(\gamma_{e})=\left\{
\begin{array}{lll}
C_{2}\gamma_{e}^{-p},&& \gamma_{m}<\gamma_{e}<\gamma_{c},\\
\frac{1+Y(\gamma_{c})}{1+Y(\gamma_{e})}C_{2}\gamma_{c}\gamma_{e}^{-p-1},&&\gamma_{c}<\gamma_{e},
\end{array}\right.
\end{equation}
where $\gamma_{m}$ is the minimum injection electron Lorentz factor, and defining $\gamma_{c}$ as the Lorentz factor above which electrons are cooled efficiently over the age of the system. $C_{1}$ and $C_{2}$ are constants. The resulting SED and light curves of the SSC components can then be obtained using this electron distribution. The approximate analytical forms of the SSC spectra in different spectral regimes are obtained in Nakar et al. (2009). It is clearly seen that the KN effect affects the SSC spectrum significantly. In the present paper, we calculate the synchrotron and SSC spectra numerically, taking into account the KN effect.

We now calculate the broadband SEDs at $t=100$ s and $t=10$ h after the burst, and the light curves at 100 GeV for the two types of media. The derived model parameter include isotropic kinetic energy ($E_{\rm k, iso}$), the energy partition factors of the electrons ($\epsilon_e$), the magnetic field ($\epsilon_B$), the initial Lorentz factor of the fireball ($\Gamma_0$) , $p$, $n_{0}$, and $A_{\ast}$.
We use the following reference parameter values: $E_{\rm k, iso}=1 \times10^{53}\ \rm erg$, $\epsilon_e = 0.3$, $\epsilon_B = 1\times10^{-4}$, $p = 2.4$, $\Gamma_0=300$, and $z=0.4$. The number densities of the external medium are taken as $n_{0}=1$ cm$^{-3}$ or $n_{0}=0.1$ cm$^{-3}$ for the homogenous medium case, and $A_{\ast}=1$ or $A_{\ast}=0.1$ for the wind medium case, respectively.

\begin{figure}[htbp]
\centering
\includegraphics[width=0.4\textwidth, angle=0]{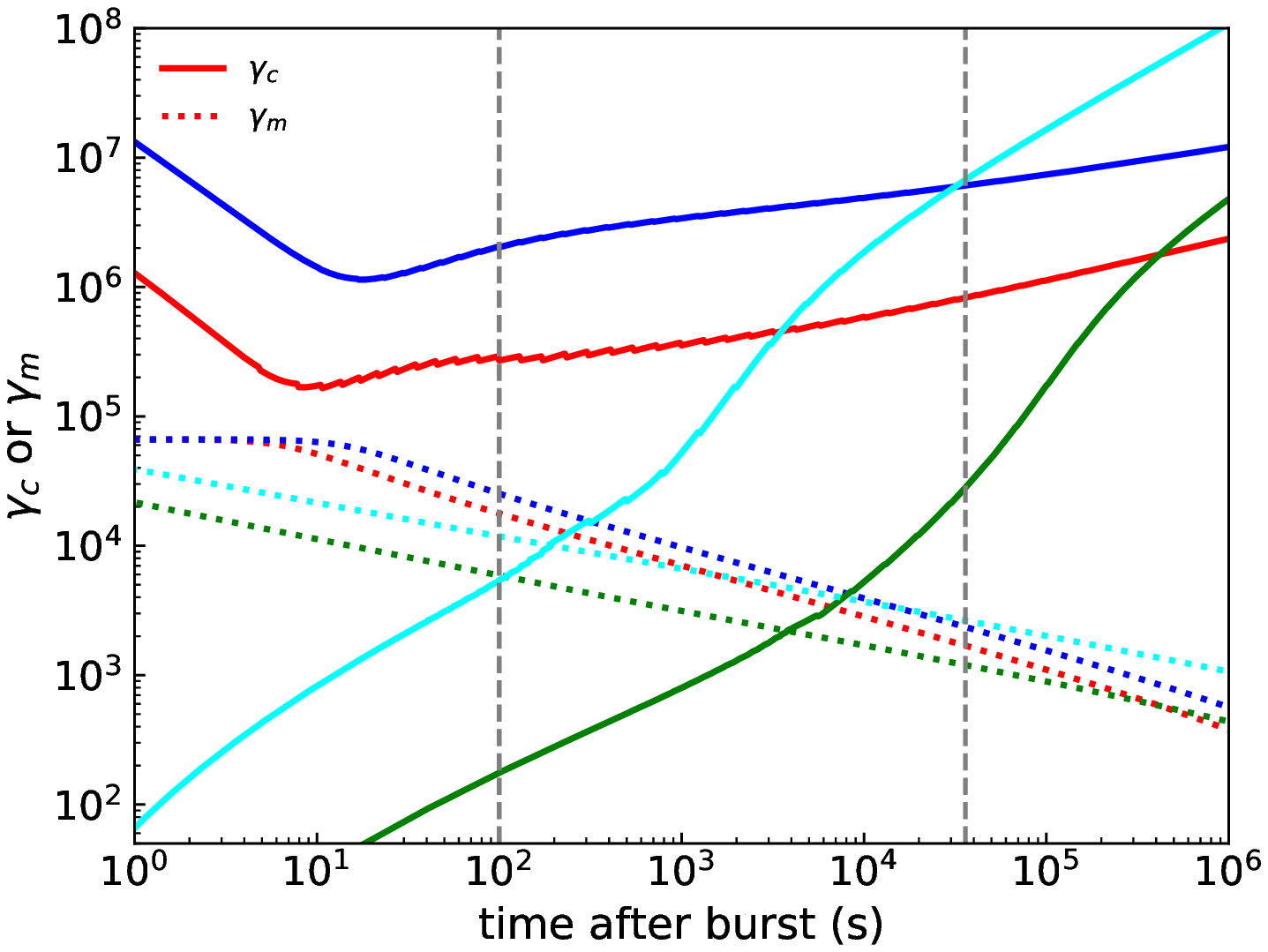}
\includegraphics[width=0.4\textwidth, angle=0]{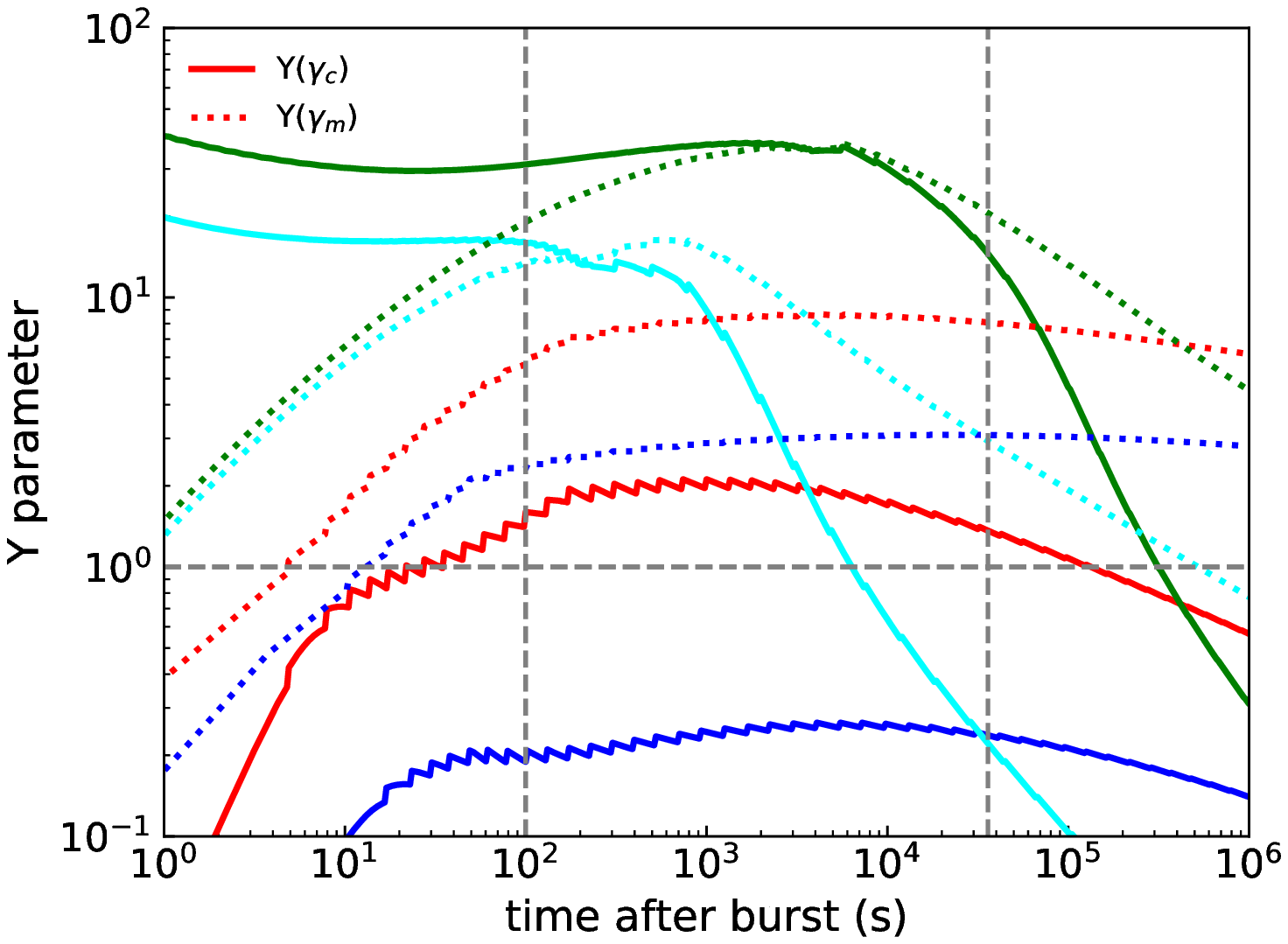}
\caption{Left panel: The values of $\gamma_{c}$ and $\gamma_{m}$ as a function of time. The solid line and dashed line represent the values of $\gamma_{c}$ and $\gamma_{m}$, respectively. The red and blue lines represent the case of homogenous medium with $n=1$ cm$^{-3}$ and $n=0.1$ cm$^{-3}$, respectively. The green and cyan lines represent the case of wind medium with  $A_{\ast}=1$ and $A_{\ast}=0.1$, respectively. Other parameter values used are: $E_{\rm k, iso}=1\times10^{53}\, \rm erg$, $\epsilon_e = 0.3$, $\epsilon_B = 1\times10^{-4}$, $p = 2.4$, $\Gamma_0=300$, and $z = 0.4$. Right panel: Compton parameters $Y(\gamma_{c})$ (solid lines) and $Y(\gamma_{m})$ (dotted lines)  as a function of time. The different color lines have the same meaning as that of the left panel.}
\label{Ypara}
\end{figure}

We first show the evolution of $\gamma_c$ and $\gamma_m$ as a function of time for two types of media in Figure~\ref{Ypara} (left panel). For typical parameter values, the radiating electrons are in the slow-cooling regime in the homogenous density case, while they are in the fast-cooling in the wind medium at early time. The Compton parameters for electrons with $\gamma_c$ and $\gamma_m$ are shown in the right panel of Figure~\ref{Ypara}. For the slow-cooling case, $Y(\gamma_c)$ reflect roughly the flux ratio between the SSC component and the synchrotron component, while for the fast-cooling case, the ratio is roughly described by $Y(\gamma_m)$. In the homogenous-density medium (slow-cooling) case, a larger density leads to a larger $Y(\gamma_c)$. However, in the wind medium case, the Compton parameter (denoted by $Y(\gamma_m)$) is not sensitive to the density. In the wind medium case, the Compton parameter $Y(\gamma_m)$ increases to a value about 10 for typical parameter values at early time, and then decreases quickly with time.

\begin{figure}[htbp]
\centering
\includegraphics[width=0.4\textwidth, angle=0]{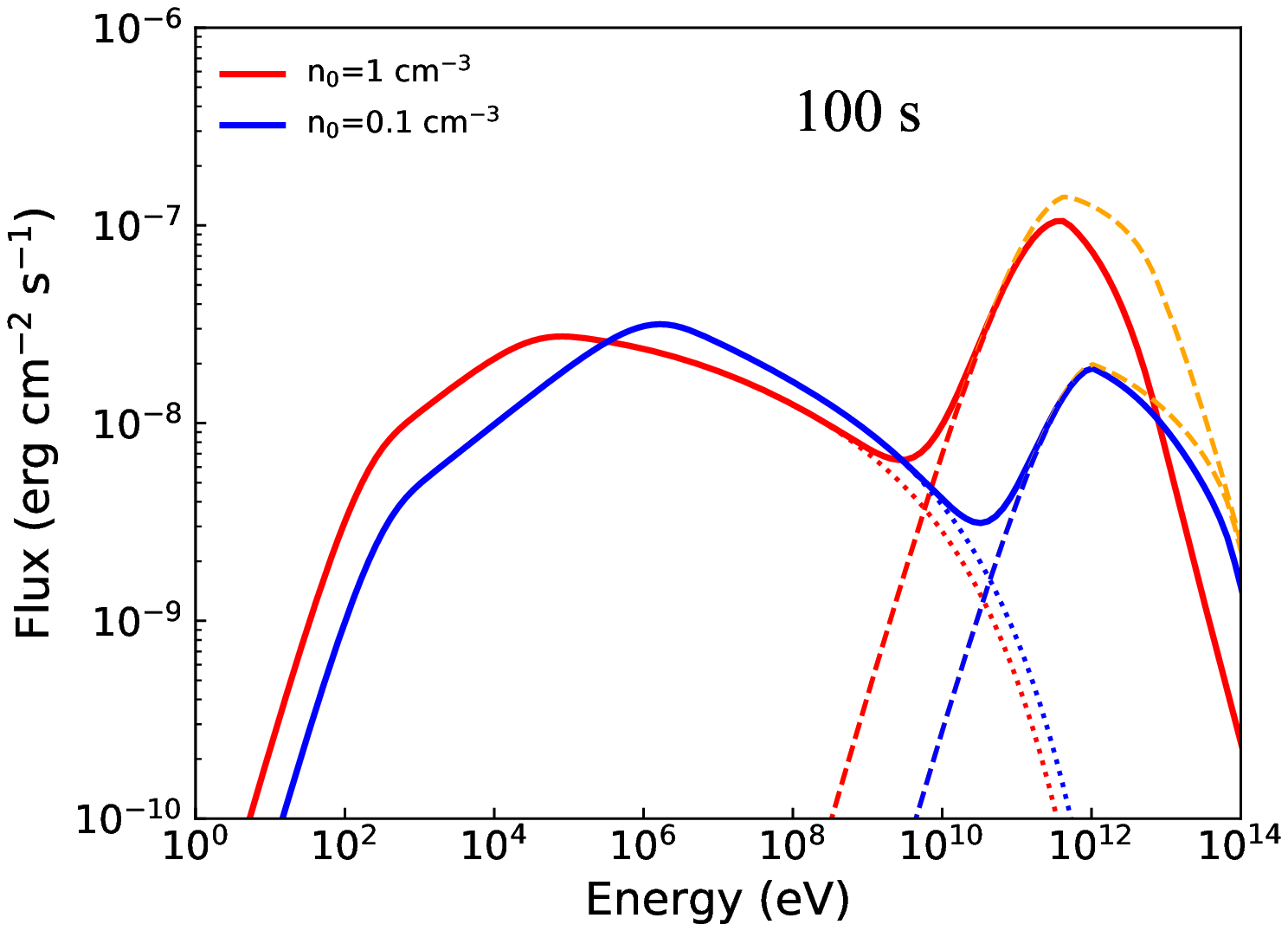}
\includegraphics[width=0.4\textwidth, angle=0]{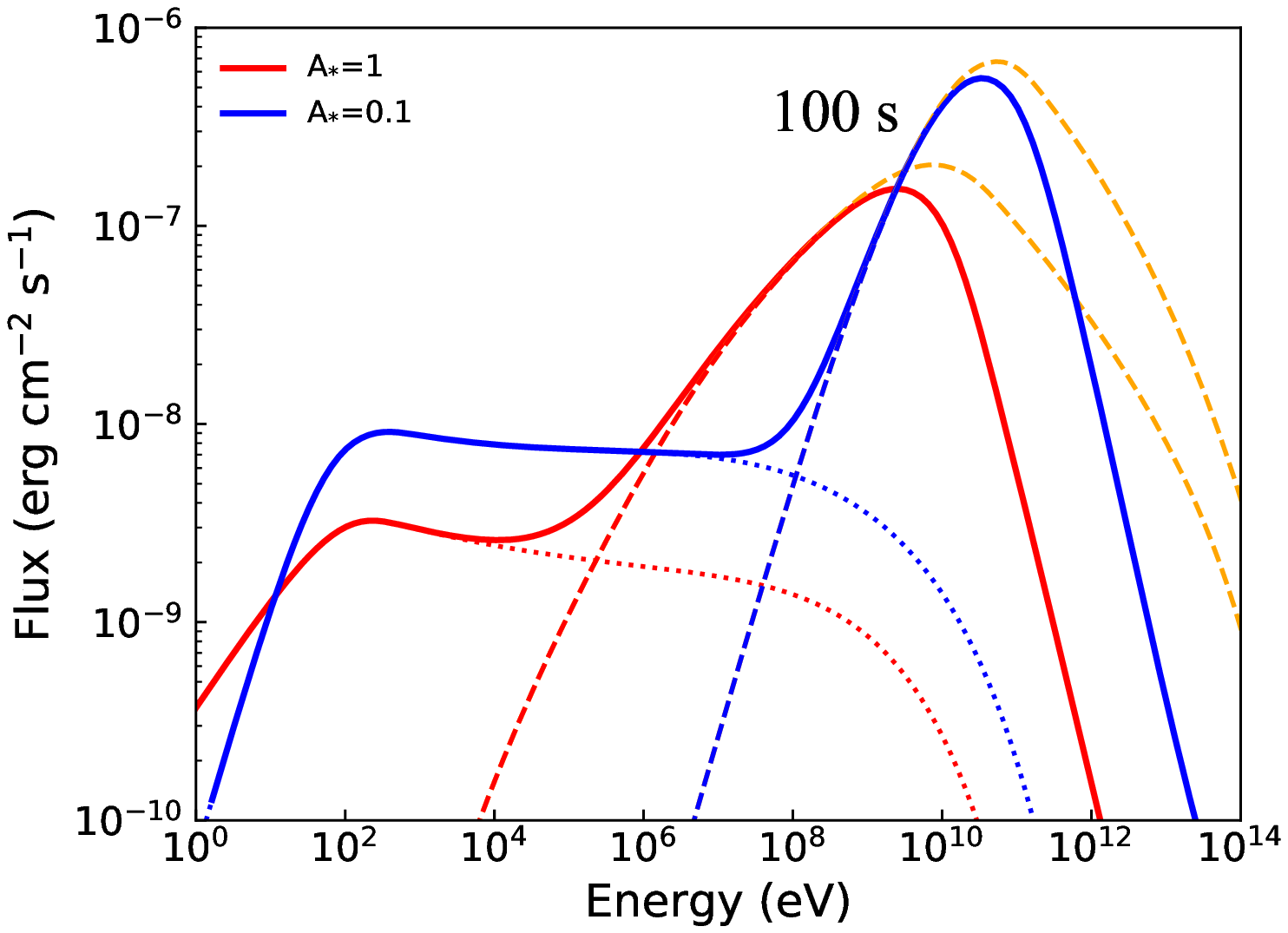}
\includegraphics[width=0.4\textwidth, angle=0]{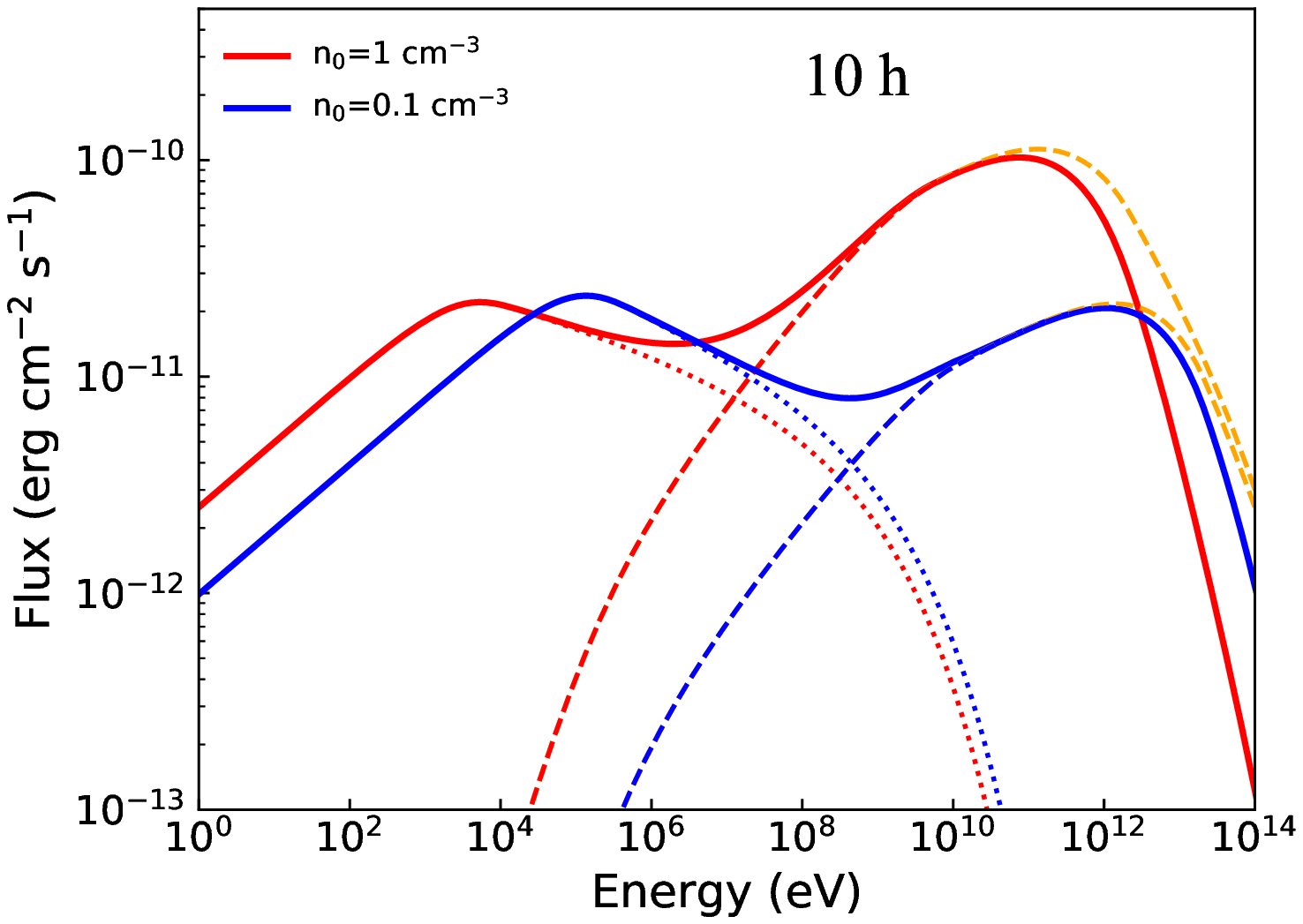}
\includegraphics[width=0.4\textwidth, angle=0]{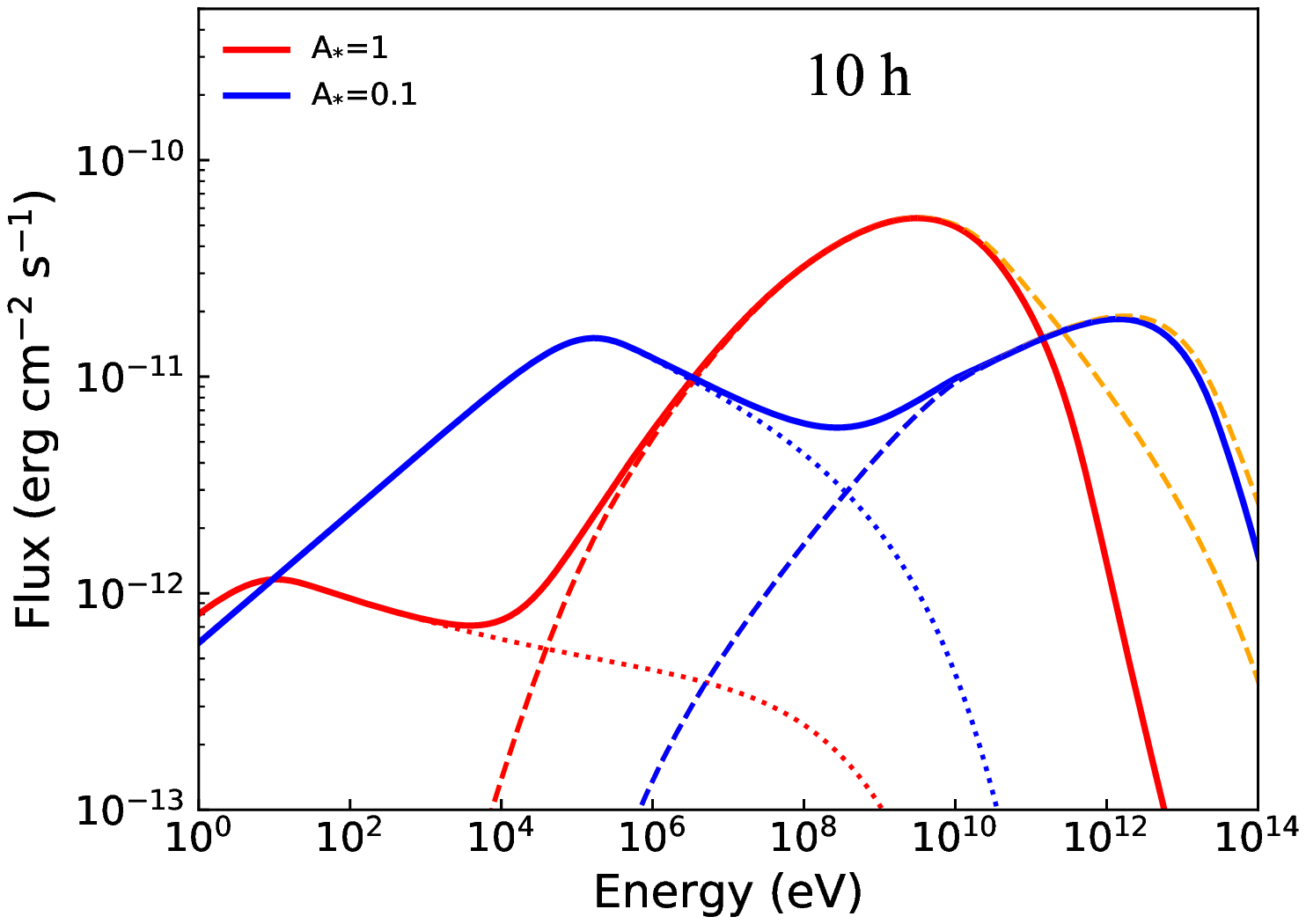}
\caption{Broadband spectral energy distributions (SEDs) of GRB afterglows in the early stage (t=100 s after the GRB trigger) and late stage (t=10 h after the GRB trigger) in the homogeneous medium with $n=1$ cm$^{-3}$ and $n=0.1$ cm$^{-3}$, and the wind medium with $A_{\ast}=1$ and $A_{\ast}=0.1$, respectively. Other parameter values used are: $E_{\rm k, iso}=1\times10^{53} \,\rm erg$, $\epsilon_e = 0.3$, $\epsilon_B = 1\times10^{-4}$, $p = 2.4$, $\Gamma_0=300$, and $z = 0.4$. The yellow dashed lines represent the emission from the SSC process without considering the absorption in the source. The solid lines represent the sum of the emission from the synchrotron radiation (the dotted lines) and the absorbed radiation of the SSC component (the dashed lines).}
\label{SEDs}
\end{figure}

The SEDs of the SSC and synchrotron emissions are shown in Figure~\ref{SEDs}. It can be seen that the peak energy of the SSC emission is larger for a lower circum-burst density in both the homogenous-density medium and wind medium cases. This is due to that a lower density results in a larger $\gamma_c$ and $\gamma_m$ in both cases. The flux ratio between the SSC component and the synchrotron component follows the evolution of $Y(\gamma_c)$ for the slow-cooling case and and $Y(\gamma_m)$ for the fast-cooling case. The ratio decreases more quickly with time in the wind medium case than that in the homogenous-density medium case. These features can be used to distinguish the two types of media.

\begin{figure}[htbp]
\centering
\includegraphics[width=0.4\textwidth, angle=0]{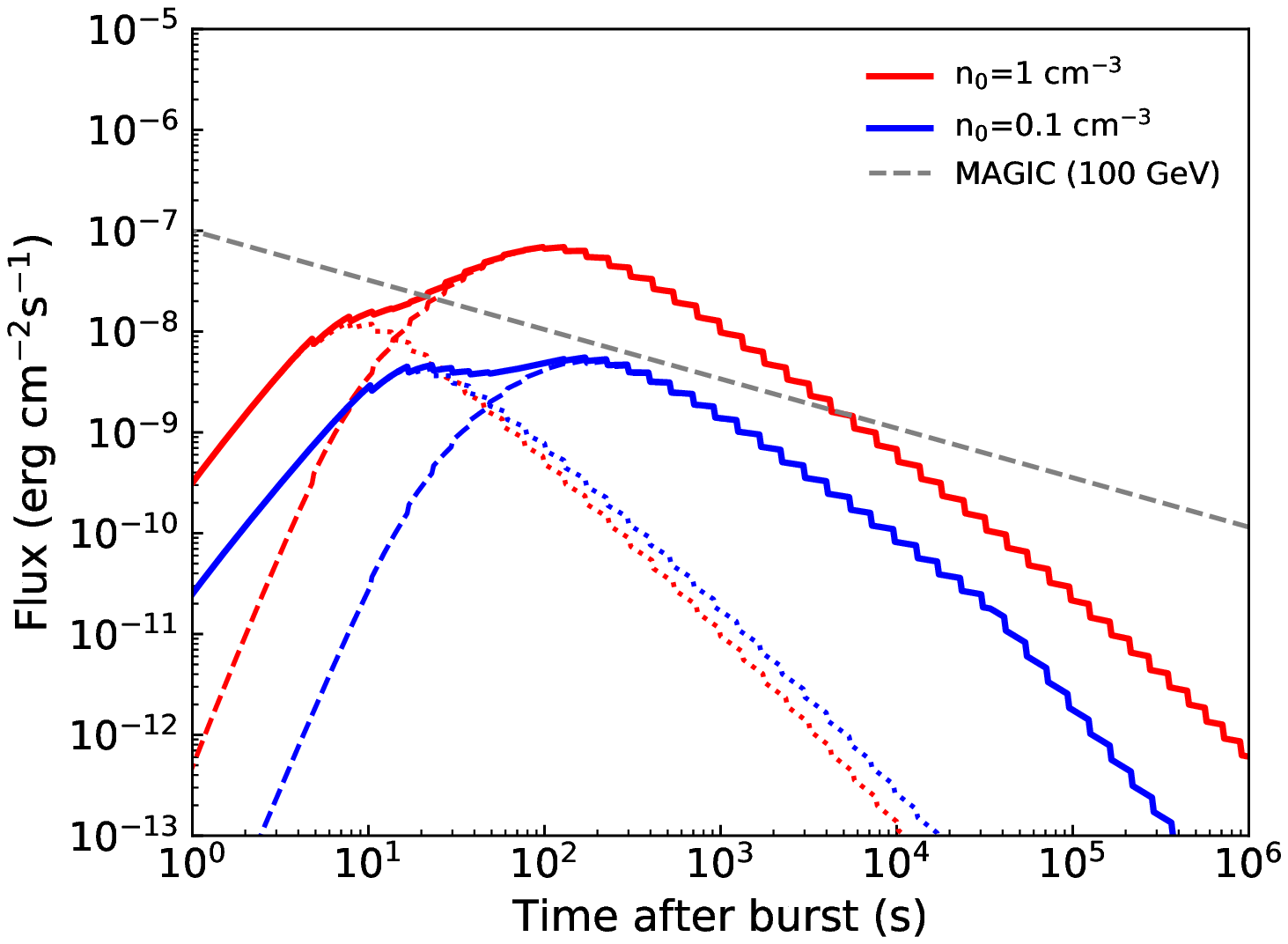}
\includegraphics[width=0.4\textwidth, angle=0]{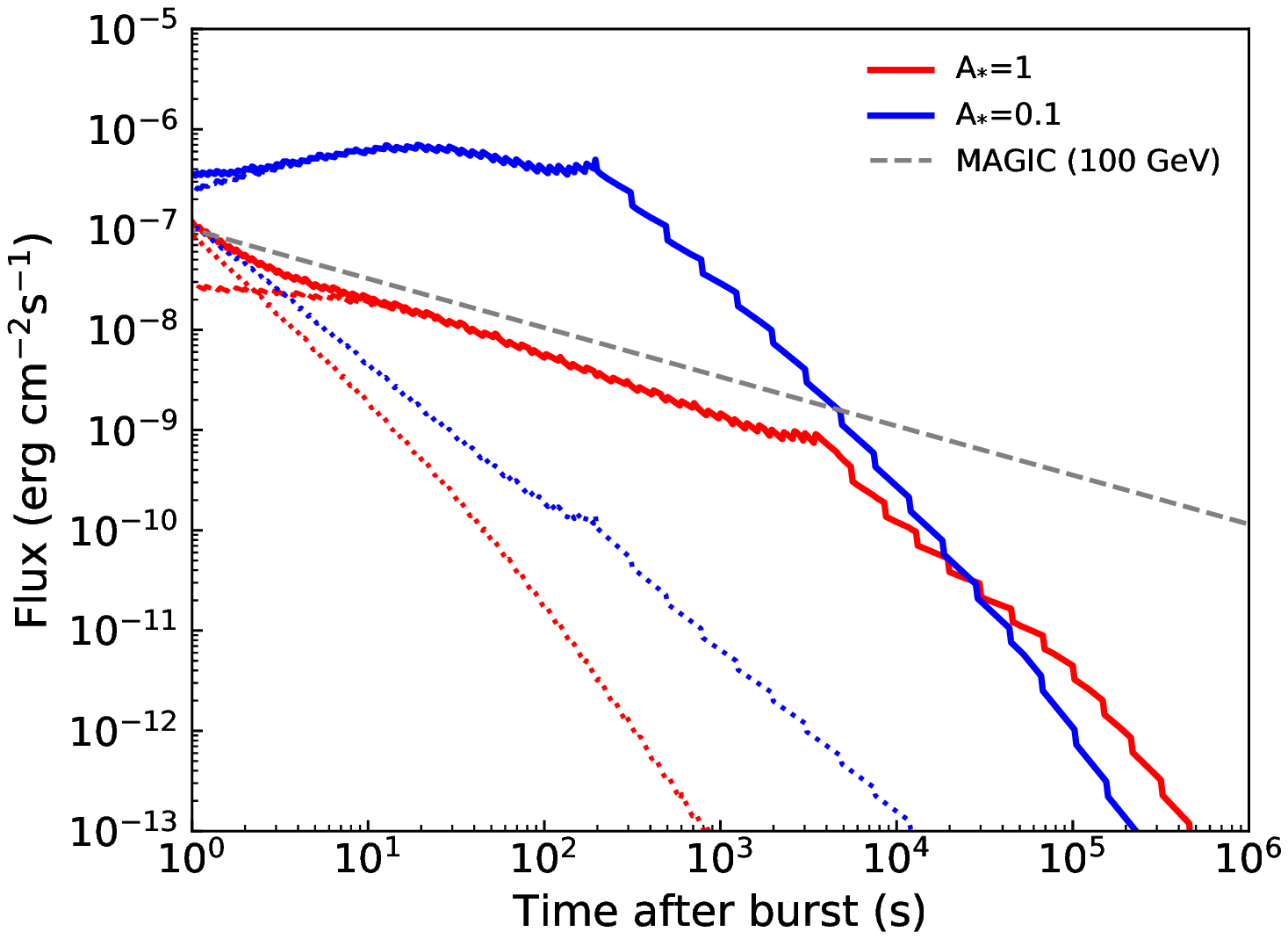}
\caption{Light curves of GRB afterglows at 100 GeV  compared with the sensitivity of MAGIC telescope in the homogeneous medium with $n=1$ cm$^{-3}$ and $n=0.1$ cm$^{-3}$ (left panel), and the wind medium  with  $A_{\ast}=1$ and $A_{\ast}=0.1$ (right panel), respectively. The grey dashed line represent the sensitivity curve of MAGIC at 100 GeV (Takahashi et al. 2008). The solid lines represent the sum of the emission from the synchrotron radiation (the dotted lines) and the absorbed radiation of the SSC component (the dashed lines). The used parameter values are the same as  that in  Figure~\ref{Ypara}. }
\label{LCs}
\end{figure}

Figure~\ref{LCs} illustrates light curves at 100 GeV in the homogeneous density (left panel) and wind (right panel) medium cases. In the homogeneous density case, the SSC emission dominates over the synchrotron emission before tens of  s after the burst. The SSC emission could be detectable by MAGIC (Major Atmospheric Gamma Imaging Cerenkov Telescope)$\footnote{https://magic.mpp.mpg.de/}$ at $t<1$ h post the GRB trigger for the case of $n=1$ cm$^{-3}$.
For the wind case, the SSC component at 100 GeV dominates over the synchrotron emission from the very beginning of the afterglow phase and the flux could be detectable up to $1$ h post the GRB trigger in the case of $A_{\ast}=0.1$. In this wind medium case, a plateau phase is clearly seen in the light curve at the early stage ($<10^{3}\rm s$), which corresponds to the analytical result of $F_{\nu}\propto (1+Y_c)^{-2} t^{0}$ in the frequency range of $\nu_{c}^{IC}<\nu<\nu_{m}^{IC}$ (see, e.g., Wang et al. (2019)).
After this, a rapid decay is seen at later time, which corresponds to $F_{\nu}\propto t^{-p}$ in the frequency range of $\nu_{m}^{IC}<\nu<\nu_{c}^{IC}$. When the wind density is too high (e.g., $A_{\ast}=1$), the flux at 100 GeV is significantly lower, mostly due to a larger internal $\gamma\gamma$ absorption in the source. Because the spectral regime of the observed frequency is different in the case of $A_{\ast}=1$, the light curve of the SSC emission at 100 GeV is also different. Generally, the evolution of light curve at 100 GeV is milder in the homogenous-density medium case than that in the wind medium case. These features can be used to distinguish the two types of media. The observed light curve of the sub-TeV emission from GRB 190114C agrees more with the homogenous density case, as has been modeled in some previous works (Wang et al. 2019; MAGIC Collaboration et al. 2019b, see also Fraija et al. 2019).

\section{Electromagnetic Cascade Emission of the  absorbed TeV photons}
\subsection{The Cascade Process}
\begin{figure}[htbp]
\centering
\includegraphics[width=0.5\textwidth, angle=0]{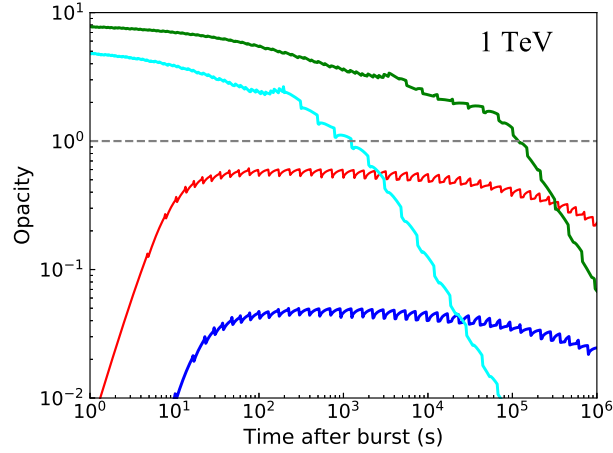}
\caption{Opacity of a  photon with energy of 1 TeV in the source as a function of time. The red and blue lines represent the case of homogenous medium with $n=1$ cm$^{-3}$ and $n=0.1$ cm$^{-3}$, respectively. On the other hand, the green and cyan lines represent the case of wind medium with $A_{\ast}=1$ and $A_{\ast}=0.1$, respectively. The grey dash line represent $\tau_{\gamma\gamma}=1$.}
\label{Opacity}
\end{figure}

The high energy photons with energy of $\rm \varepsilon_{\gamma}$ suffer from pair-production absorption by interacting with target photons with energy $\epsilon_{t}\ga \Gamma^{2}(m_{e}c^{2})^{2}/\rm \epsilon_{\gamma}$ in the source. Then a cascade process is induced, and the energy of high energy photons is redistributed into lower energy photons, until the opacity of secondary photons becomes $\tau_{\gamma\gamma}<1$. As an example, Figure~\ref{Opacity} shows the opacity of a photon with energy of 1 $\rm TeV$ as a function of time\footnote{We do not consider the EBL absorbtion in the intergalactic space in our calculation.}. For the reference parameter values, the opacity in the homogeneous medium case is $\tau_{\gamma\gamma}<1$ for 1 $\rm TeV$ photons from the beginning of the afterglow. However, the opacity is $\tau_{\gamma\gamma}>1$ in the denser wind medium. This is due to that a denser medium leads to a lower bulk Lorentz factor of the forward shock and hence a higher opacity for high-energy photons. Below we perform a comparative study of the cascade emission between the homogeneous and wind media.

Following B$\ddot{\rm o}$ttcher et al. (2013), we adopt a semi-analytical method to calculate the cascade emission for the purpose of an efficient calculation of cascades. We assume the high-energy photons of the SSC and synchrotron radiation as the first-generation photon field, which are derived from the primary electron spectra (i.e., Eq.2 and Eq.3). The injection rate of the first-generation photons is denoted by $\dot{N}_{\epsilon}^{0}$.
Then the secondary high-energy photons are produced through synchrotron emission and IC processes, whose production rate is denoted by $\dot{N}_{\epsilon}^{\rm sec}$.
Considering the absorption, the spectrum of escaping (observable) photons can
be calculated as
\begin{equation}
\begin{split}
\dot{N}_{\epsilon}^{\rm esc}=(\dot{N}_{\epsilon}^{0}+\dot{N}_{\epsilon}^{\rm sec})(\frac{1-e^{-\tau_{\gamma\gamma}(\epsilon)}}{\tau_{\gamma\gamma}(\epsilon)}),
\end{split}
\end{equation}
where $\tau_{\gamma\gamma}(\epsilon)$ is the optical depth of photons due to $\gamma\gamma$ absorption.

We then calculate the production rate of electrons/positron pairs due to $\gamma\gamma$ absorption. In the $\gamma\gamma$ absorption of
a high-energy photon of energy $\epsilon$, one of the produced particles will take the major fraction, $f_{\gamma_{e}}$, of the photon energy.
Hence, an electron/positron pair with energies $\gamma_1=f_{\gamma_{e}} \epsilon$ and $\gamma_2=(1-f_{\gamma_{e}}) \epsilon$ is produced. Following B$\ddot{\rm o}$ttcher et al. (2013), we adopt $f_{\gamma_{e}}=0.9$ in our calculation.
Defining an absorption factor $f_{\rm abs}(\epsilon)$  as
\begin{equation}
\begin{split}
f_{\rm abs}(\epsilon)\equiv1-\frac{1-e^{-\tau_{\gamma\gamma}(\epsilon)}}{\tau_{\gamma\gamma}(\epsilon)},
\end{split}
\end{equation}
the pair production rate can be written as
\begin{equation}
\begin{split}
\dot{N}_{\rm e}^{\gamma\gamma}(\gamma_{e})=f_{\rm abs}(\epsilon_{1})(\dot{N}_{\epsilon_{1}}^{0}+\dot{N}_{\epsilon_{1}}^{\rm sec})+
f_{\rm abs}(\epsilon_{2})(\dot{N}_{\epsilon_{2}}^{0}+\dot{N}_{\epsilon_{2}}^{\rm sec}),
\end{split}
\end{equation}
where $\epsilon_{1}=\gamma_{e}/f_{\gamma_{e}}$ and $\epsilon_{2}=\gamma_{e}/(1-f_{\gamma_{e}})$ (B\"{o}ttcher et al. 2013; Veres et al. 2017).

The energy loss of electrons through synchrotron and SSC  processes is given by
\begin{equation}
\begin{split}
\dot{\gamma_{e}}=\frac{4}{3}\frac{c\,\sigma_{\rm T}}{m_{e}c^{2}}\gamma_{e}^{2}[U_{\rm B}+U_{\rm syn}f_{\rm KN}(\gamma_{e})],
\end{split}
\end{equation}
where $\sigma_{\rm T}$ is the Thomson cross section, $m_{e}$ is the electron mass, $c$ is the light speed, and $U_{\rm B}$ and $U_{\rm syn}$ are, respectively, the energy density of magnetic filed and synchrotron photons. Here $f_{\rm KN}$ is a correction factor accounting for the KN effect, i.e., $f_{\rm KN}(\gamma_{e})=\int_{\epsilon,\rm min}^{\epsilon,\rm max} f_{\rm KN}(\kappa)u(\epsilon)d \epsilon/U_{\rm syn}$, where $u(\epsilon)$ is the differential energy distribution of the synchrotron photons and $\kappa=4\gamma_{e}\epsilon$. $f_{\rm KN}$ is approximated as (Moderski et al. 2005):
\begin{equation}
f_{\rm KN}(\kappa)\simeq\left\{
\begin{array}{lll}
1 && \kappa\ll1 \,(\rm Thomson\, limit)\\
\frac{9}{2\kappa^{2}}(\ln\kappa-\frac{11}{6}) && \kappa\gg1 \, (\rm KN\, limit).
\end{array}\right.
\end{equation}

In the calculation, we divide the time interval logarithmically. To achieve sufficient accuracy, we adopt a very small time increment $\delta t=(10^{0.01}-1)t$ in the numerical calculation. The calculation of distribution of electron in the cascade at time $t+\delta t$ can be divided into two parts. The first part is the cascade electrons accumulated from the beginning to time $t$, the other part is the electrons newly produced in time from time $t$ to $t+\delta t$. For the accumulated electrons, the cooling effect can be included by considering the electron number conservation $N_{\rm e}^{\rm cool}(\gamma_{e},t+\delta t)d\gamma_{e}=N_{\rm e}^{\rm sec}(\gamma_{e}^{\ast},t)d\gamma_{e}^{\ast}$, where $\gamma_{e}^{\ast}$ is the electron Lorentz factor at time t, and due to the cooling effect, the Lorentz factor will decrease from $\gamma_{e}^{\ast}$ to $\gamma_{e}$ during the time interval $\delta t$. For the newly produced electrons in time interval $\delta t$, the calculation is divided into two cases according to the relation between the cooling timescale of electrons and the time interval.
When the electron cooling timescale is less than the time interval (i.e., $t_{e}^{\rm cool}(\gamma_{e})<\delta t$), the cascade process tends to be in a quasi-steady state and the electron distribution is given by (B$\ddot{\rm o}$ttcher et al. 2013),
\begin{equation}
\begin{split}
\begin{aligned}
N_{\rm e}^{\rm sec}(\gamma_{e}&,t+\delta t)=N_{\rm e}^{\rm sec}(\gamma_{e}^{\ast},t)\frac{d\gamma_{e}^{\ast}}{d\gamma_{e}}\\
&+\frac{1}{\dot{\gamma}_{e}}\int_{\gamma_{e}}^{\infty} d\widetilde{\gamma_{e}}\,
\dot{N}_{\rm e}^{\gamma\gamma}(\widetilde{\gamma_{e}},t+\delta t).
\end{aligned}
\end{split}
\end{equation}
While for the case of $t_{e}^{\rm cool}(\gamma_{e})>\delta t$, the electron distribution is given by
\begin{equation}
\begin{split}
\begin{aligned}
N_{\rm e}^{\rm sec}(\gamma_{e},t+\delta t)=&N_{\rm e}^{\rm sec}(\gamma_{e}^{\ast},t)\frac{d\gamma_{e}^{\ast}}{d\gamma_{e}}\\
&+\dot{N}_{\rm e}^{\gamma\gamma}(\gamma_{e},t+\delta t)\,\delta t.
\end{aligned}
\end{split}
\end{equation}

\subsection{The Cascade Emission in homogeneous and Wind Media}
\begin{figure}[htbp]
\centering
\includegraphics[width=0.4\textwidth, angle=0]{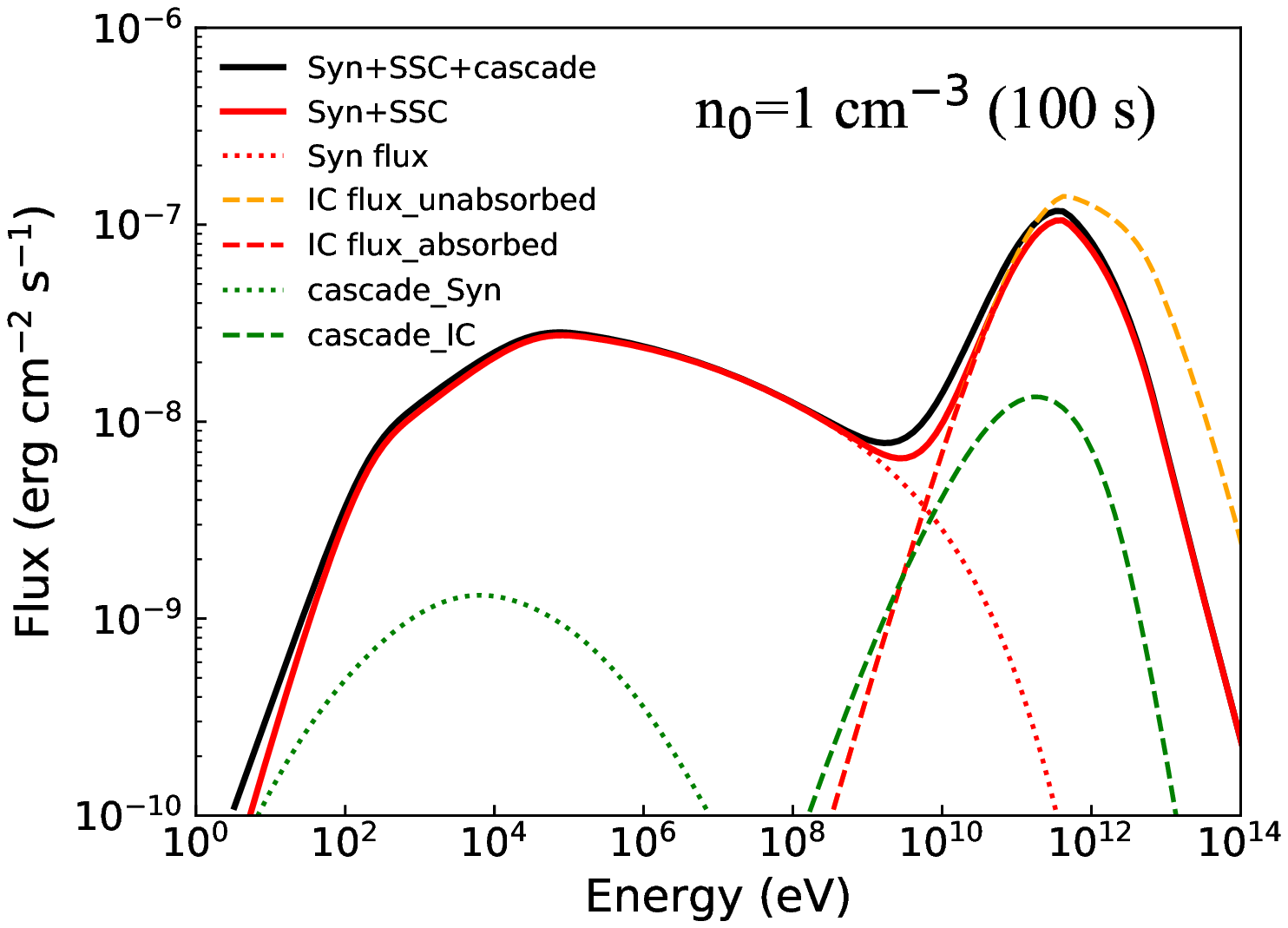}
\includegraphics[width=0.4\textwidth, angle=0]{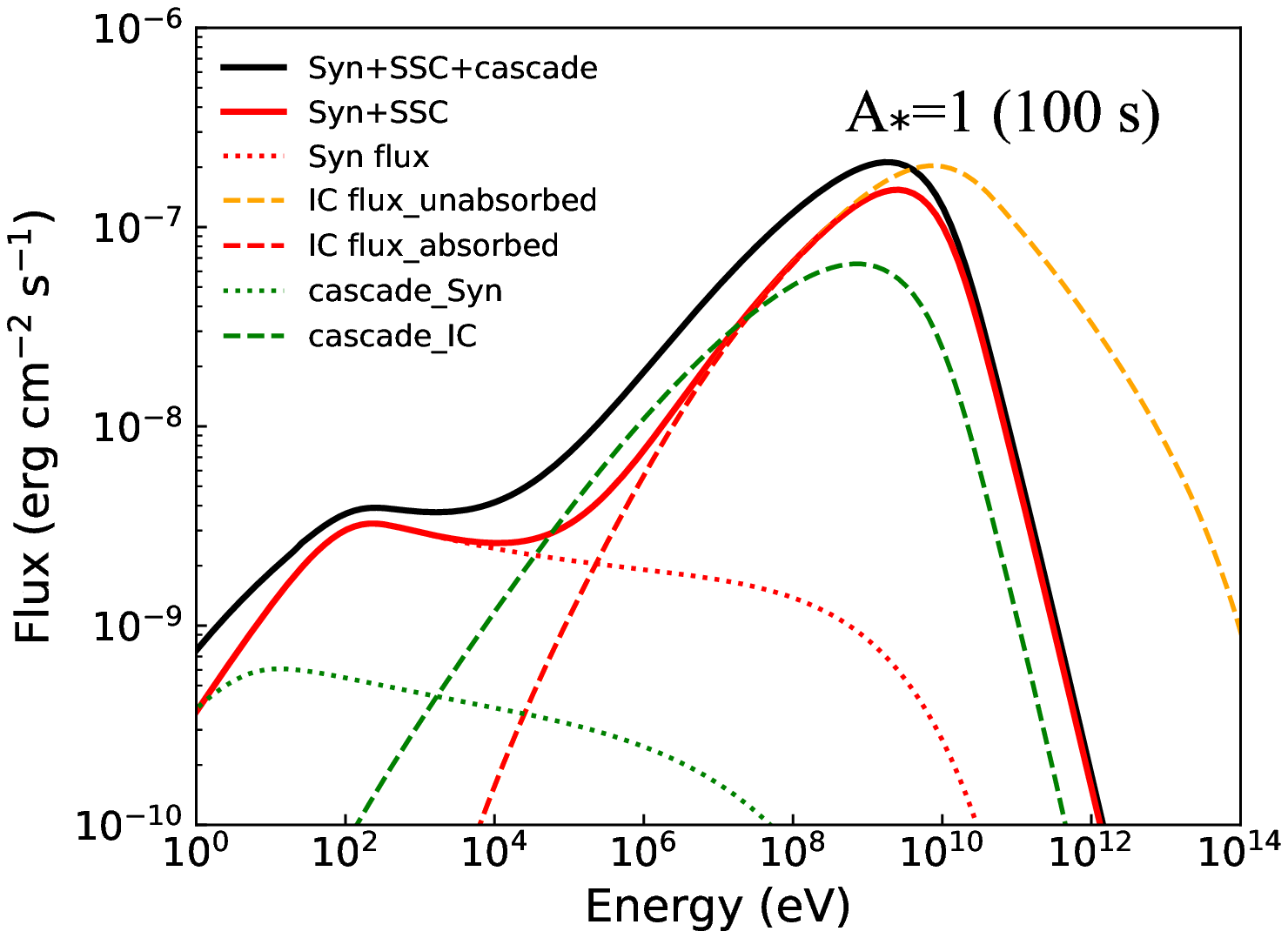}
\includegraphics[width=0.4\textwidth, angle=0]{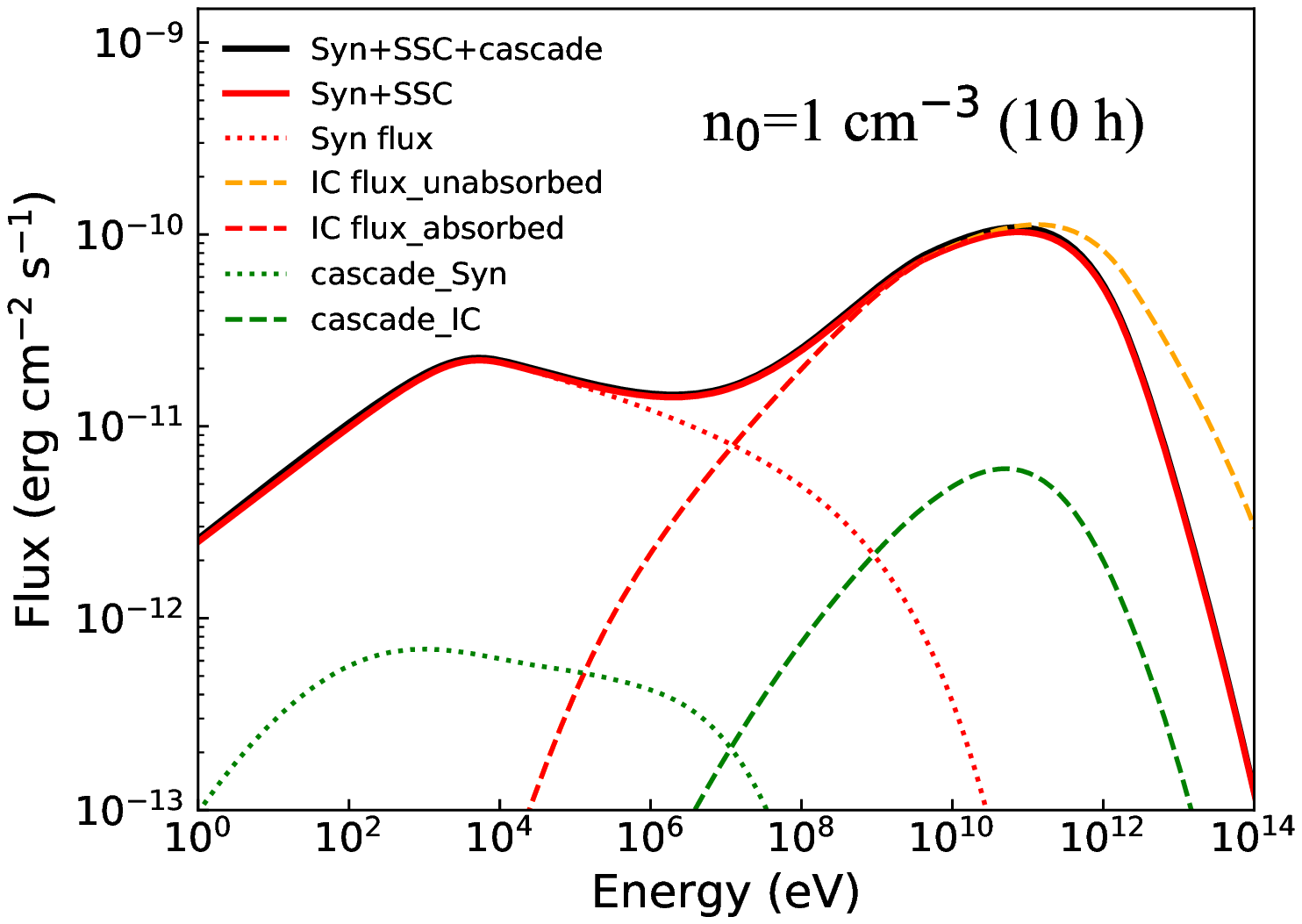}
\includegraphics[width=0.4\textwidth, angle=0]{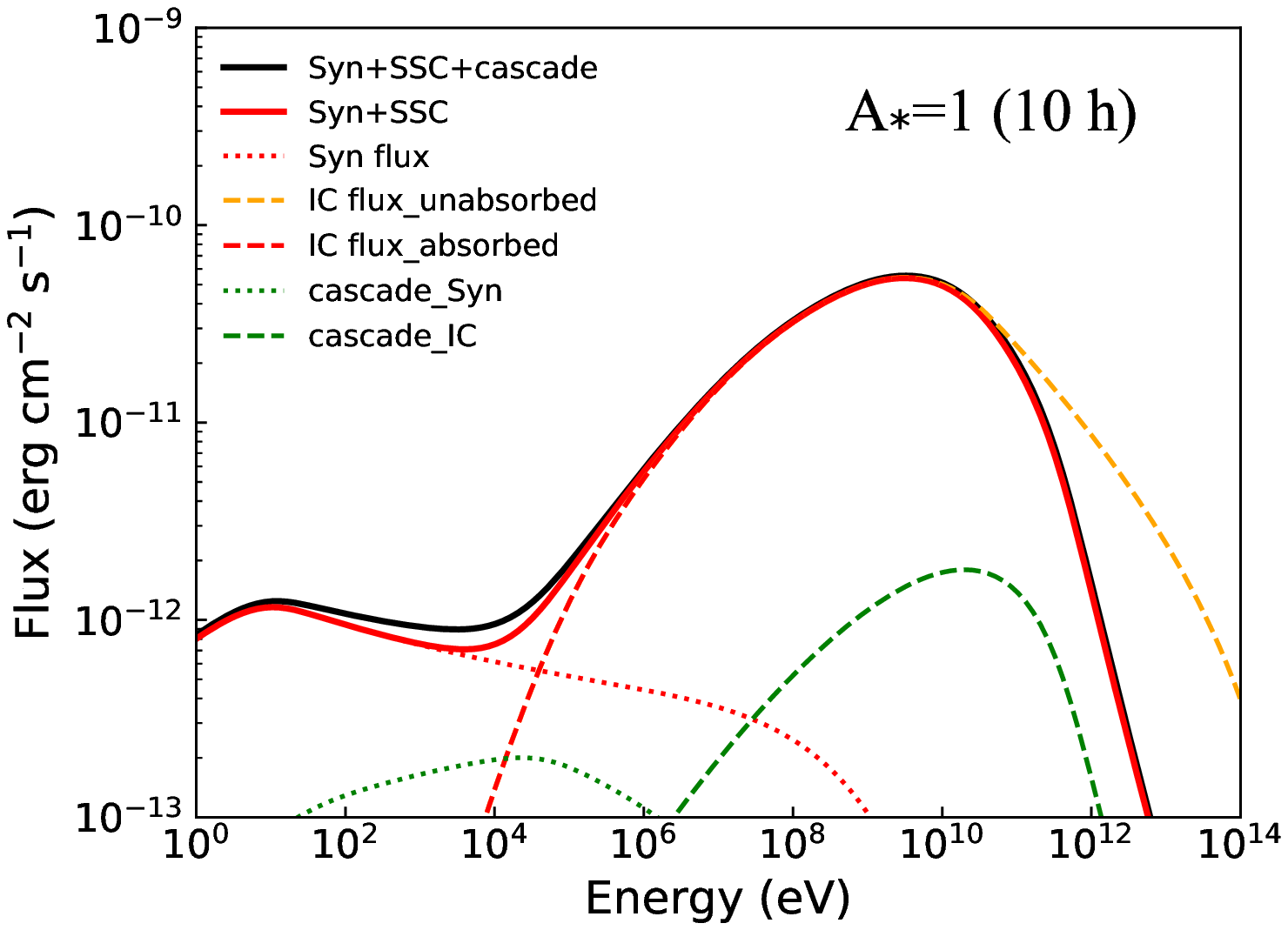}
\caption{SEDs of GRB afterglows in the early (t=100 s, $top$ panels) and late (t=10 h, $bottom$ panels) stages in the homogeneous medium with $n=1$ cm$^{-3}$ ($left$ panels), and the wind medium with $A_{\ast}=1$ ($right$ panels), respectively. The solid black lines represent the sum of the emission from the synchrotron radiation (the dotted red lines),  SSC process (the dashed red lines), and the cascade radiation (the green lines). The red solid lines represent the sum of the emission from the synchrotron radiation and the absorbed radiations of the SSC process only. The yellow dashed line represents the emission from the SSC process without considering the $\gamma\gamma$ absorbed effect. The green dotted and dashed lines represent the cascade emission from the synchrotron radiation and SSC radiation of the secondary electrons, respectively. The parameter values used are the same as that in Figure~\ref{Ypara}. }
\label{Cascade_SED}
\end{figure}

Taking the cascade emission into account, we re-calculate the SEDs and the light curves of the afterglows with the same parameter set as mentioned above. Figure~\ref{Cascade_SED} illustrates the broadband SEDs of the afterglows at $t=100$ s and $t=10$ h in homogeneous and wind media with number densities of $n_{0}=1$ cm$^{-3}$ and $A_{\ast}=1$, respectively.

For the case of $n_{0}=1$ cm$^{-3}$ at $t=100$ s (the upper left panel of Figure~\ref{Cascade_SED}), the cascade synchrotron component at $\sim {\rm 1 eV}$  and the cascade SSC emission at $\sim 100$ GeV are both comparable to that of the primary electron population (the cascade emission is marked as black solid line). In the cases of $n_{0}=1$ cm$^{-3}$ at $t=10{\rm h}$, the broadband SEDs are overwhelmingly dominated by the radiations of the primary syn+SSC electron population. In the wind medium, the cascade emission is also sensitive to $A_{\ast}$. The cascade radiations contribute significantly to the whole SED for $A_{\ast}=1$ at the early stage ($t=100$ s). The cascade SSC emission contributes significantly to the left shoulder of the SSC bump. On the other hand, the cascade synchrotron emission  dominates the optical flux ($\sim 1$ eV). At late epoch of $t=10$ h, the cascade emission only contributes weakly to the SED around $\sim 10^4$ eV, and is ignorable in other bands in comparison with the emission from the primary electron population.

\begin{figure}[htbp]
\centering
\includegraphics[width=0.35\textwidth, angle=0]{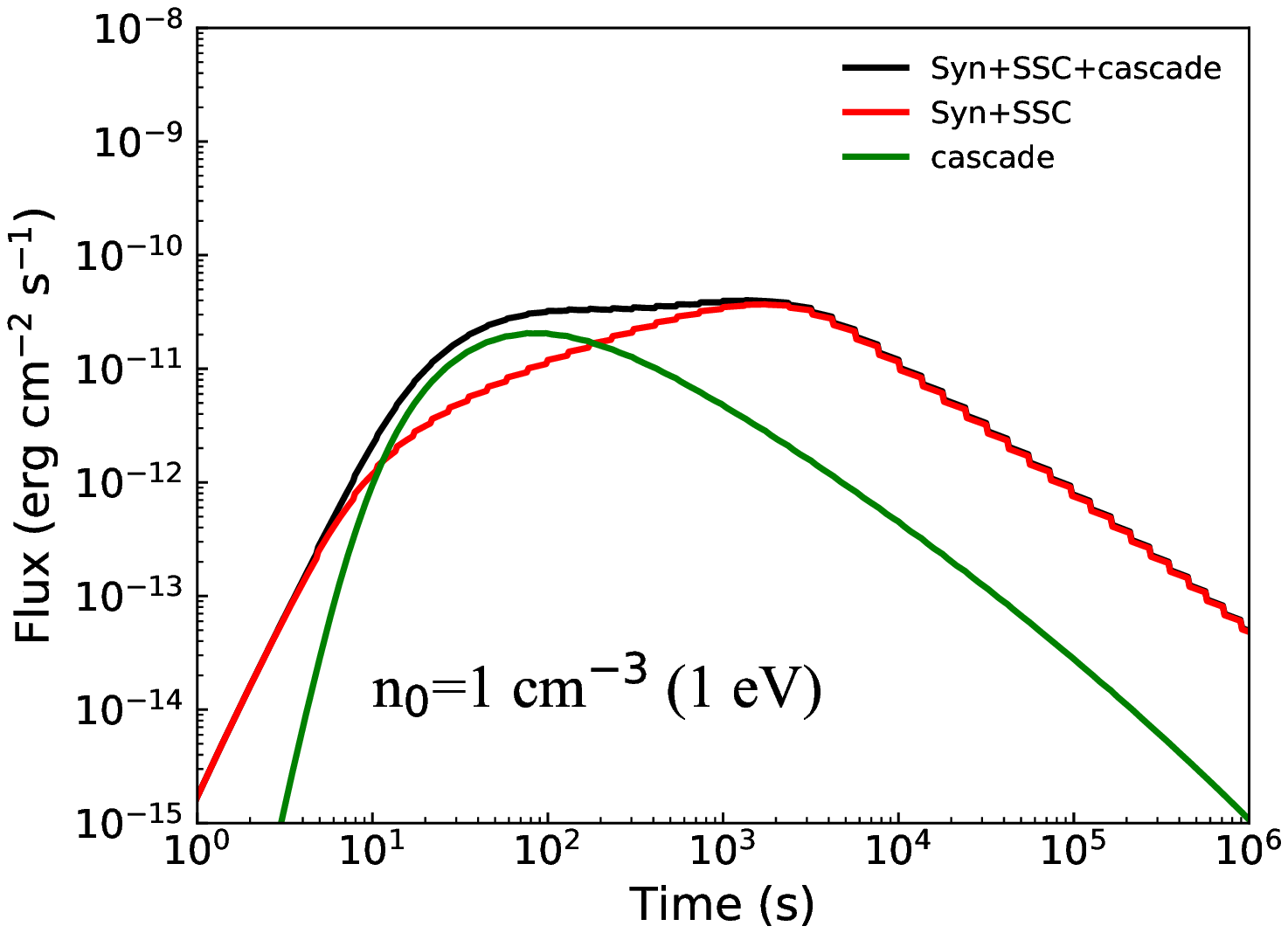}
\includegraphics[width=0.35\textwidth, angle=0]{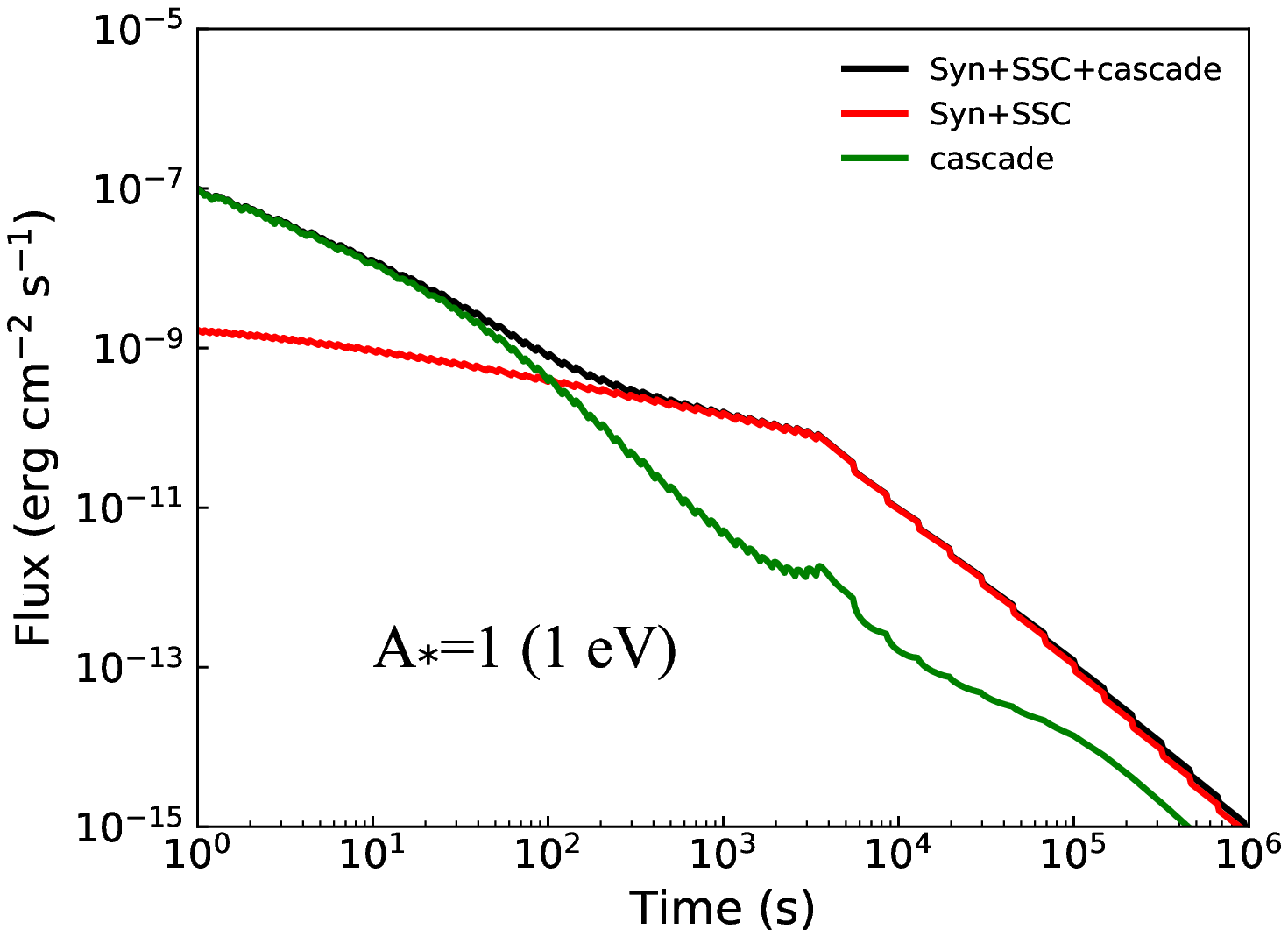}
\includegraphics[width=0.35\textwidth, angle=0]{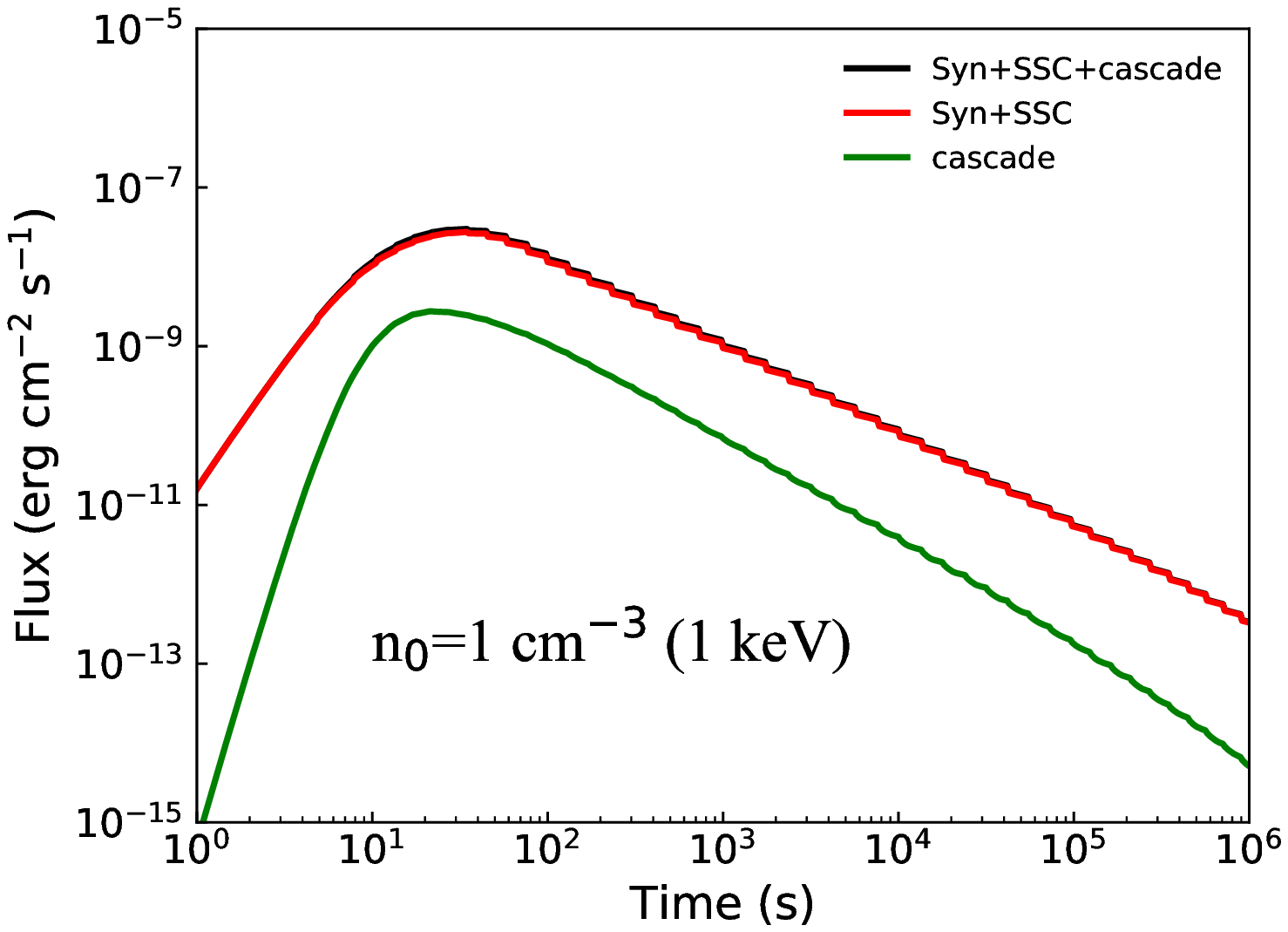}
\includegraphics[width=0.35\textwidth, angle=0]{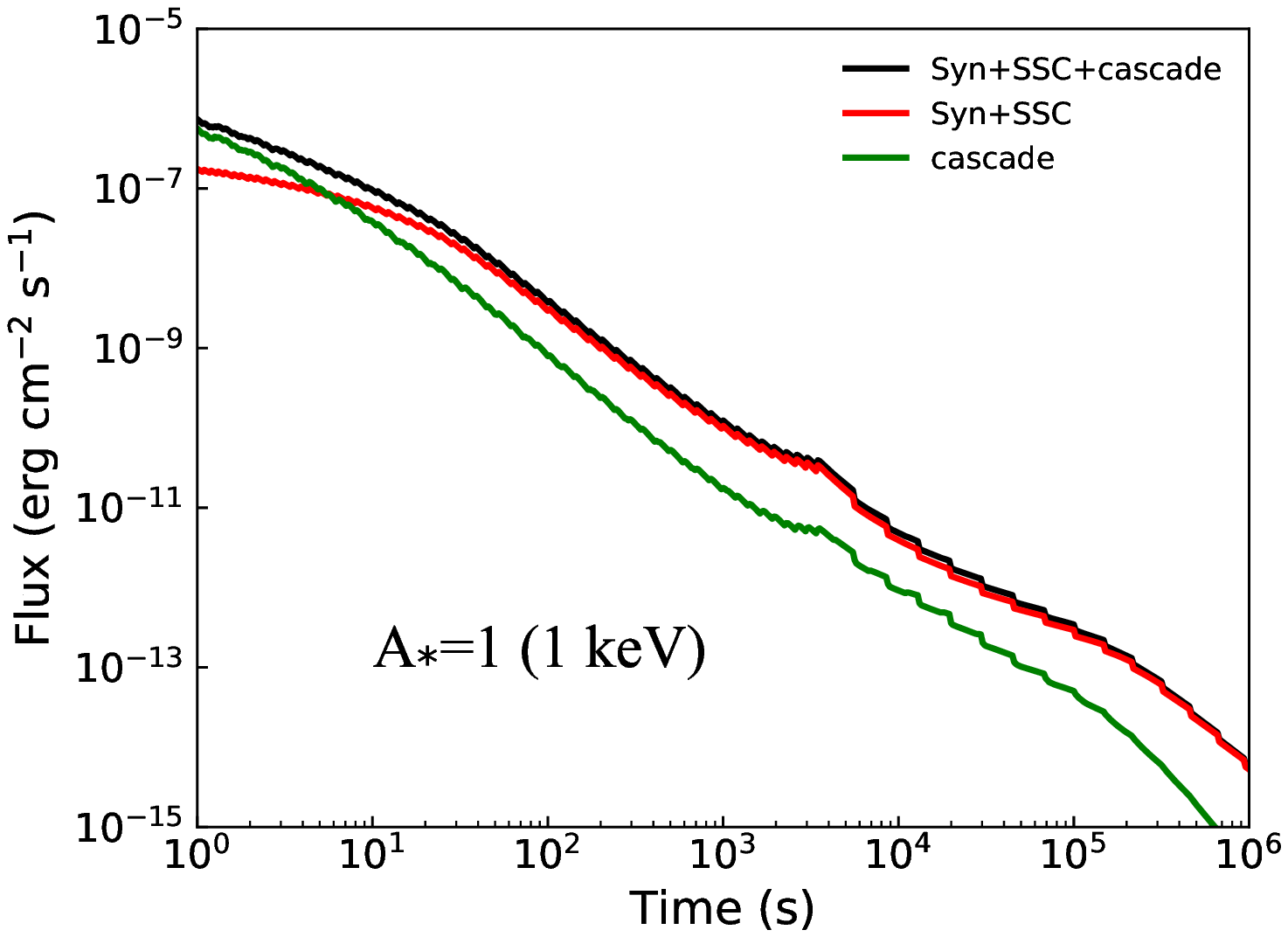}
\includegraphics[width=0.35\textwidth, angle=0]{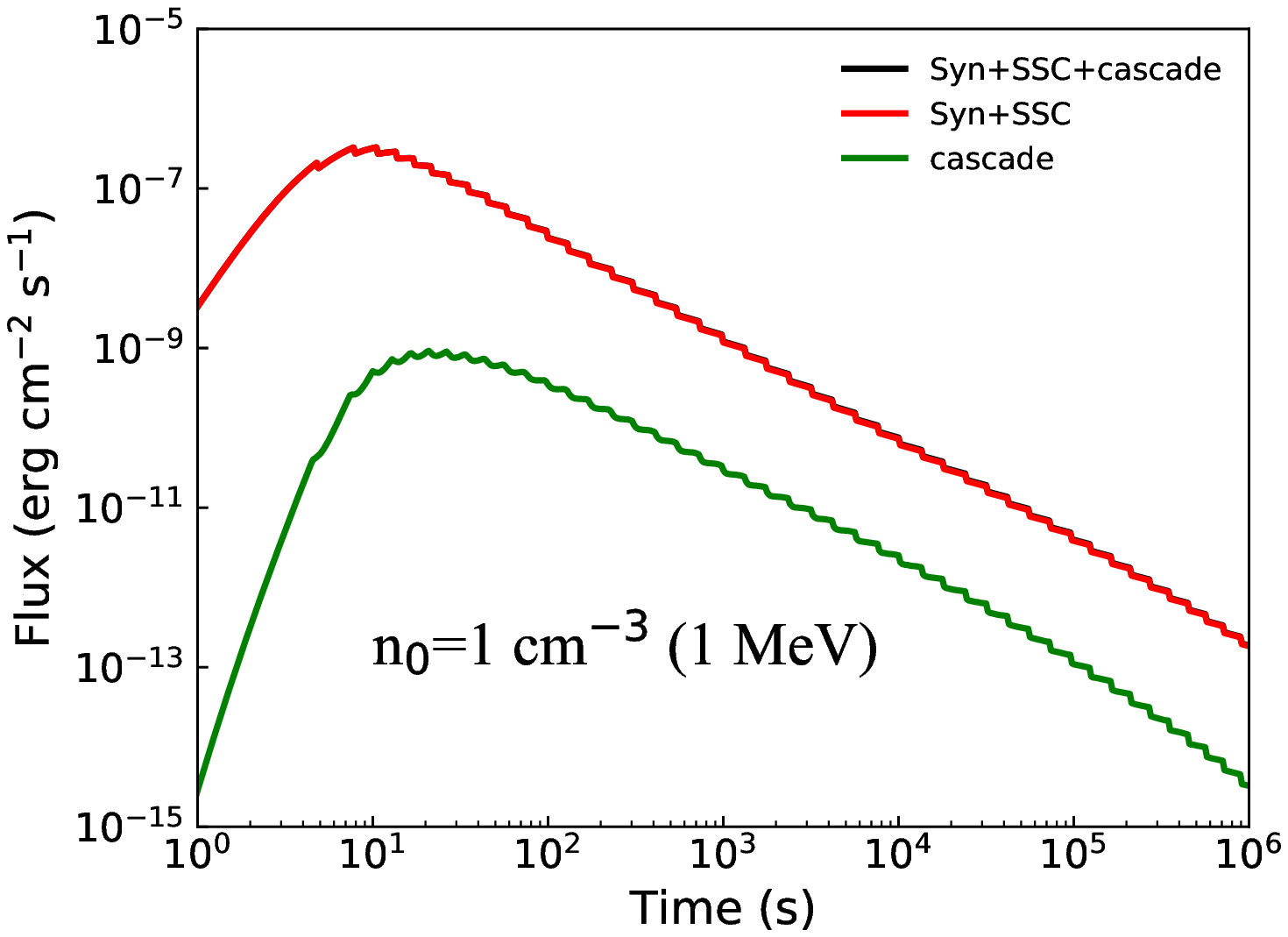}
\includegraphics[width=0.35\textwidth, angle=0]{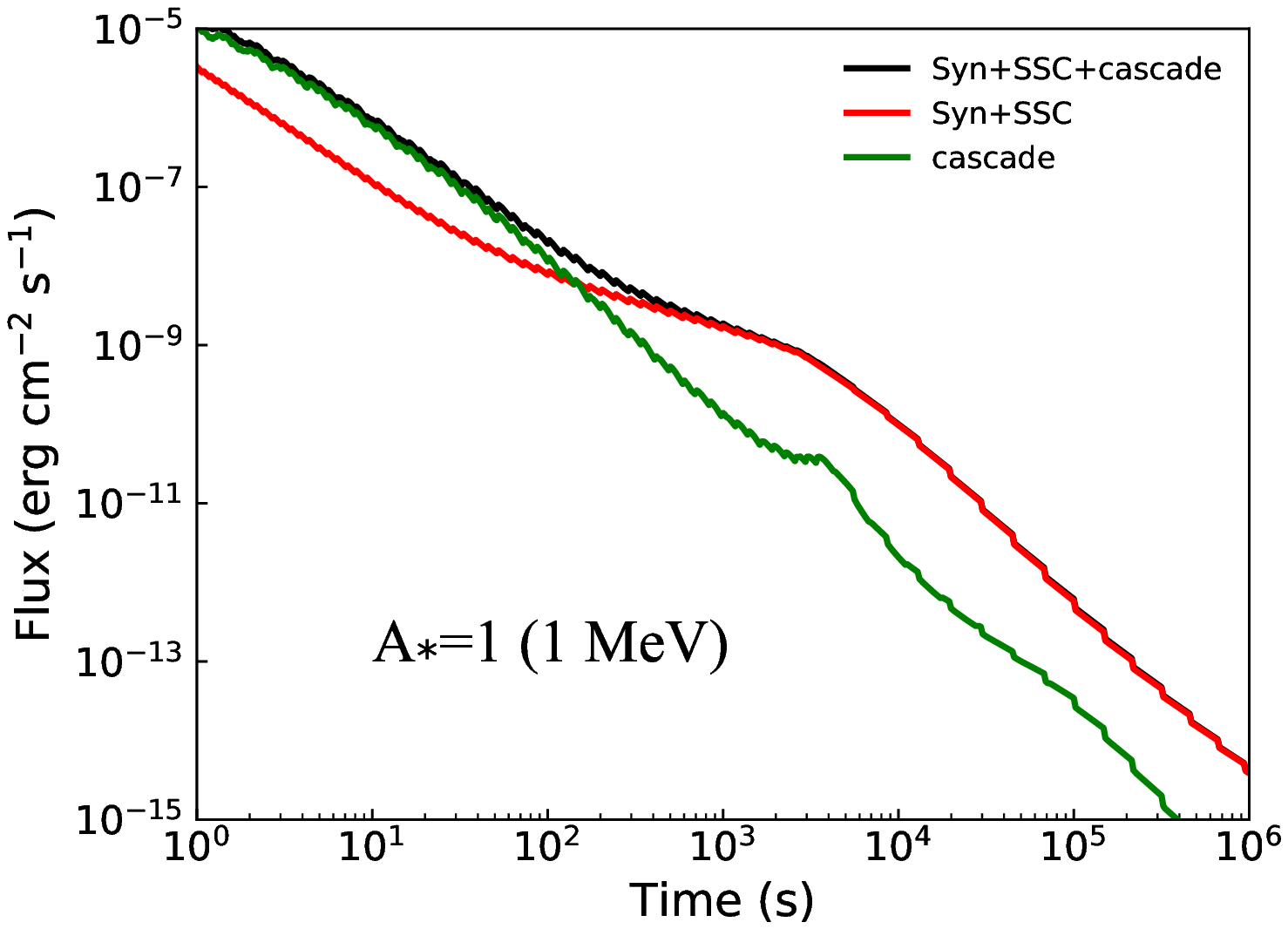}
\includegraphics[width=0.35\textwidth, angle=0]{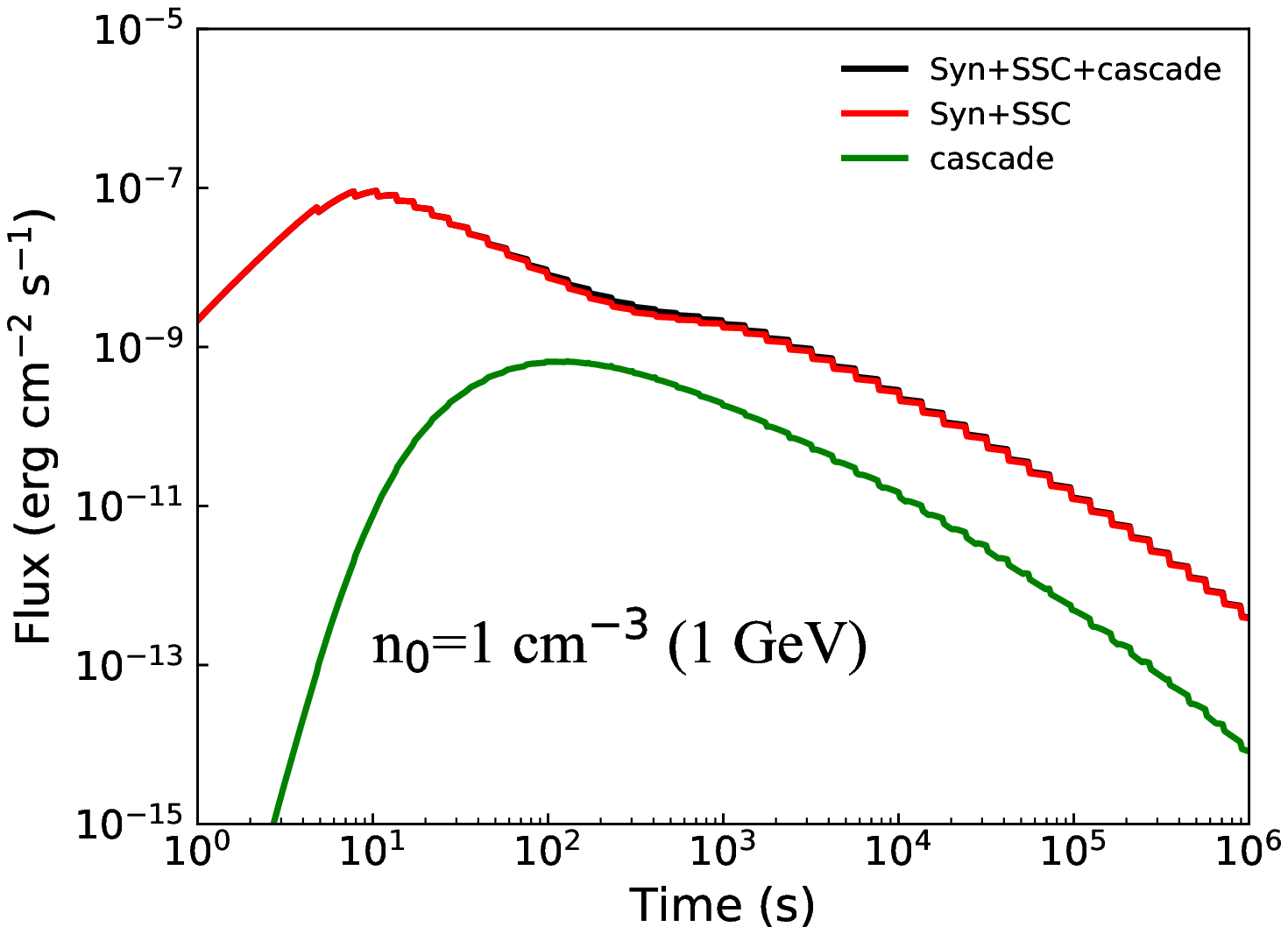}
\includegraphics[width=0.35\textwidth, angle=0]{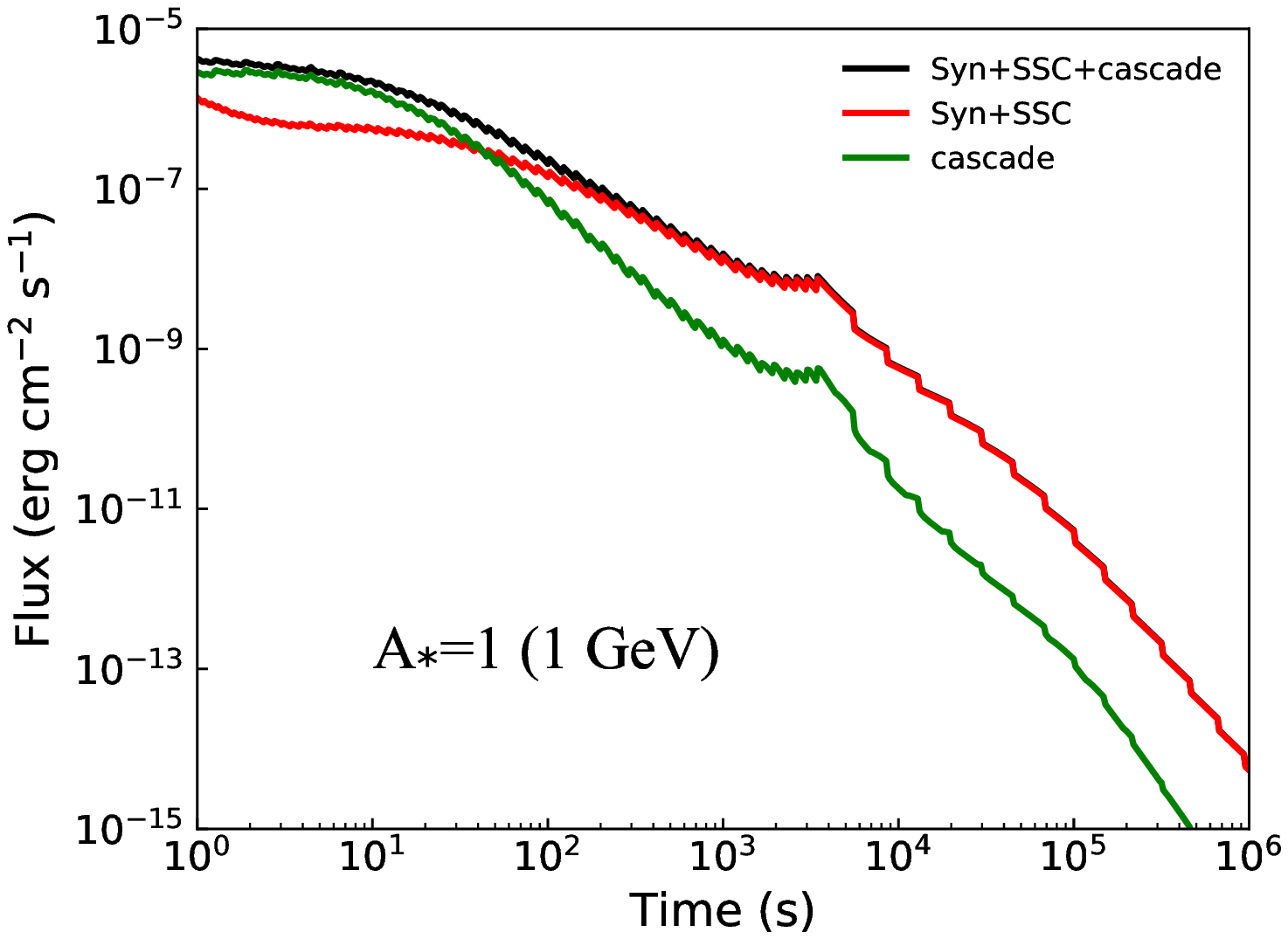}
\caption{Light curves of GRB afterglows in several energy bands (1eV, 1keV, 1MeV, and 1GeV) in the homogeneous medium with number density of $n=1$ cm$^{-3}$ ($left$ panels) and the wind medium with $A_{\ast}=1$ ($right$ panels).  The black solid lines represent the sum of the emission from synchrotron emission, SSC emission, and the cascade emission. The red solid lines represent the sum of the emission from the synchrotron emission and the SSC emission after considering the $\gamma\gamma$ absorption. The green solid lines represent the sum of the cascade emission from the synchrotron emission and SSC emission of the secondary electrons produced in the pair production. The parameter values used are the same as that in Figure~\ref{Ypara}. Note that the spikes in some plots arise from the discontinuity of KN factors, which are obtained approximately in our calculation,  in the transition between the fast cooling and slow cooling cases.}
\label{Cascade_LC}
\end{figure}

Figure~\ref{Cascade_LC} shows {the corresponding mono-frequency light curves} in homogeneous and wind media with number densities of $n_{0}=1$ cm$^{-3}$ and $A_{\ast}=1$, respectively.
For the homogeneous medium, the extra cascade emission component shows up mainly in the optical ($\sim 1$ eV) light curves. The superimposed effect of the emission from both the primary and cascade electrons flattens  the light curves at the early stage. This might explain the plateau  seen in the early optical afterglows of some GRBs (e.g., Panaitescu \& Vestrand 2011; Liang et al. 2013). The cascade emission contribute subdominantly to the X-ray afterglow for typical parameter values. It may  lead to a plateau in X-rays at early time if the density of circum-burst medium is sufficiently high.  Some GRBs also display a plateau in X-rays at late time (Fraija et al. 2020; Fraija et al. 2019a), which is, however, hard to explain with the cascade emission.  The mono-frequency light curves in wind medium are different from that  in the homogeneous medium. The light curve behaviors depend on the competition between the primary and cascade radiations. At the early stages ($t\lesssim 1000$ s), the light curves are dominated by the cascade emission, while it is  dominated by the primary SSC emission at late stage. The overlapped effect of the primary and cascade radiations makes the light curves complicated, but they generally illustrate as a shallow decay followed by a steep decay segment.

\section{Conclusions and Discussions}
We have presented a comparative analysis of the sub-TeV emission of GRB afterglows in the homogeneous and wind media in the framework of synchrotron and SSC emissions of electrons accelerated in the forward shock. The attenuation of very high energy photons in the source due to $\gamma\gamma$ absorption and the KN effect on the SSC spectrum are considered.  We find  that the flux of the  SSC emission  could be detectable with current Imaging Atmospheric Cherenkov Telescopes (IACT) up to $\sim 10$ h post the GRB trigger for GRB 190114C-like bright GRBs in an ISM medium with number density $n\sim 1$ cm$^{-3}$ or in the wind medium with $A_{\ast}\sim 0.1$. Generally, the SSC emission is stronger in the denser environment. But a too dense medium, e.g. a wind medium with $A_{\ast}= 1$, will suppress the sub-TeV emission due to the severe $\gamma\gamma$ absorption.  For future telescopes such as Cherenkov Telescope Array (CTA), the detection rate of sub-TeV emission from GRBs would be increased significantly. The light curves of the sub-TeV emission are different for the two types of media, which can be used to distinguish the circum-burst medium in the future.

The absorbed high-energy photons lead to cascade emission at low energies. In the homogenous ISM scenario, the cascade emission could be comparable to the synchrotron of the primary electrons in the optical band and flatten the early optical afterglow light curve ($t<1$ h). In the wind medium, the cascade emission at early time is comparable or even larger than the emission of the primary electrons in a wide range of frequencies. It has been found that the observed diversity of the early optical light curves is hard to explain in the simple external shock model (e.g., Wang et al. 2015). The cascade mission  might be helpful to explain this diverse behavior of the optical afterglows, as well as X-ray afterglows. A detailed study of this possibility is, however, beyond the scope of the present paper.

\section{Acknowledgement}
This work is supported by the NSFC under the Grants No. 11625312 and No. 11851304, by the National Key R\&D program of China under the Grant No. 2018YFA0404203; by the National Natural Science Foundation of China (Grant No.11533003, 11851304, and U1731239), by the Guangxi Science Foundation and special funding for Guangxi distinguished professors (2017AD22006).

\clearpage
\newpage

\newpage

\newpage

\newpage

\end{document}